\documentclass[12pt]{article}
\usepackage{epsf,amsfonts,amssymb,epsfig,amsmath}
\usepackage{color}
\renewcommand{\text}[1]{#1}

\newcommand{\be}{\begin{equation}}
\newcommand{\ee}{\end{equation}}
\newcommand{\ben}{\begin{displaymath}}
\newcommand{\een}{\end{displaymath}}
\newcommand{\bea}{\begin{eqnarray}}
\newcommand{\eea}{\end{eqnarray}}
\newcommand{\bean}{\begin{eqnarray*}}
\newcommand{\eean}{\end{eqnarray*}}
\newcommand{\nn}{\nonumber \\}
\newcommand{\ba}{\begin{array}}
\newcommand{\ea}{\end{array}}
\newcommand{\bi}{\begin{itemize}}
\newcommand{\ei}{\end{itemize}}

\newcommand{\reef}[1]{(\ref{#1})}

\newcommand{\bbR}{{\mathbb{R}}}
\newcommand{\bbF}{{\mathbb{F}}}

\newcommand{\ft}[2]{{\textstyle\frac{#1}{#2}}}

\begin{document}

\makeatletter
\renewcommand{\theequation}{\thesection.\arabic{equation}}
\@addtoreset{equation}{section}
\makeatother

\baselineskip 18pt

\begin{titlepage}

\vfill

\begin{flushright}
Imperial/TP/2010/JG/01\\
AEI-2010-044
\end{flushright}

\vfill

\begin{center}
   \baselineskip=16pt
   \begin{Large}\textbf{
       Universal Kaluza--Klein reductions of type IIB \\*[5pt]to $N=4$ supergravity in five dimensions  }
   \end{Large}
   \vskip 1.5cm
    Jerome P. Gauntlett$^1$ and  Oscar Varela$^2$\\
   \vskip .6cm
     \begin{small}
  \textit{$^1$Theoretical Physics Group, Blackett Laboratory, \\
        Imperial College, London SW7 2AZ, U.K.}
        %E-mail: j.gauntlett, d.waldram@imperial.ac.uk}
        \end{small}\\*[.4cm]
   %     and\\*[.4cm]
      \begin{small}
      \textit{$^2$ AEI, Max-Planck-Institut f\"ur Gravitationsphysik, \\
	Am M\"uhlenberg 1, D-14476 Potsdam, Germany }
        %E-mail: j.gauntlett, d.waldram@imperial.ac.uk}
        \end{small}
   \end{center}

\vfill

\begin{center}
\textbf{Abstract}
\end{center}

\begin{quote}
We construct explicit consistent Kaluza--Klein reductions of type IIB supergravity
on $HK_4\times S^1$, where $HK_4$ is an arbitrary four-dimensional hyper-K\"ahler manifold, and
on $SE_5$, an arbitrary five-dimensional Sasaki--Einstein manifold.
In the former case we obtain the bosonic action of
$D=5$ $N=4$ (ungauged) supergravity coupled to two vector multiplets.
For the $SE_5$ case we extend a known reduction, which leads to minimal $D=5$ $N=2$ gauged
supergravity, to also include a multiplet of massive fields, containing the breathing mode
of the $SE_5$. We show that the resulting $D=5$ action is also consistent with
$N=4$ gauged supergravity coupled to two vector multiplets.
This theory has a supersymmetric $AdS_5$
vacuum, which uplifts to the class of supersymmetric $AdS_5\times SE_5$ solutions,
that spontaneously breaks $N=4$ to $N=2$,
and also a non-supersymmetric $AdS_5$ vacuum which uplifts to
a class of solutions first found by Romans.
\end{quote}

\vfill

\end{titlepage}

\setcounter{equation}{0}

%%%%%%%%%%%%%%%%%%%%%%%%%%%%%%%%%%%%%%%%%%%%%%%%%%%%%%%%%%%%%%%%%%%%%%%
%\tableofcontents
%%%%%%%%%%%%%%%%%%%%%%%%%%%%%

\section{Introduction}

Consistent Kaluza-Klein (KK) reductions provide powerful tools to construct exact solutions of $D=10$ and $D=11$ supergravity.
For example, it has been shown, at the level of the bosonic fields, that there is a consistent KK reduction of type IIB supergravity on an arbitrary five-dimensional Sasaki-Einstein space, $SE_5$, to minimal $N=2$ gauged supergravity in $D=5$ \cite{Buchel:2006gb}.
By definition, this means that
any solution of the $D=5$ gauged supergravity can be uplifted on an arbitrary $SE_5$ space to obtain an infinite class of exact solutions of type IIB supergravity, one for each choice of $SE_5$.
In particular, the supersymmetric $AdS_5$ vacuum solution uplifts to the class of supersymmetric $AdS_5\times SE_5$ solutions which are
dual to $N=1$ SCFTs in $d=4$.
There is a similar consistent KK reduction of $D=11$ supergravity on an arbitrary seven dimensional Sasaki-Einstein space, $SE_7$, to minimal
$N=2$ gauged supergravity in $D=4$ \cite{gv}. In this case the supersymmetric $AdS_4$ vacuum solution of this theory uplifts to the class of
supersymmetric $AdS_4\times SE_7$ solutions dual to $N=2$ SCFTs in $d=3$.

These two examples form part of a more general story. For any supersymmetric solution of $D=10$ or $D=11$ supergravity
consisting of a warped product of an $AdS_{d+1}$ space with an internal manifold $M$ and fluxes preserving the symmetries of
$AdS_{d+1}$, it is expected \cite{gv}
that there is always a consistent
KK reduction on $M$ to a $D=d+1$ gauged supergravity theory where one only keeps the fields of the supermultiplet containing the metric.
In the $d$ dimensional SCFT dual to the supergravity solution, these fields are dual to the superconformal current multiplet.
In the above $SE$ examples, the bosonic field content of the $D=5,4$ minimal supergravities consist of a metric and
a single gauge-field  which are indeed dual to the energy-momentum tensor and the abelian
$R$-symmetry current of the superconformal current multiplet in the dual SCFTs.
Other examples include the $N=8$ $SO(8)$ gauged supergravity arising from the KK reduction of $D=11$ supergravity on $S^7$
 and the $N=8$ $SO(6)$ gauged supergravity arising from type IIB supergravity on $S^5$,
 where the consistency has been partially demonstrated in \cite{deWit:1986iy} and \cite{Cvetic:1999xp,Lu:1999bw,Cvetic:2000nc,
 Khavaev:1998fb} , respectively,
 and also examples discussed in \cite{Gauntlett:2006ai,Gauntlett:2007sm}.

It is worth noting that
 almost all work on consistent KK reductions, of this kind,
 works at the level of the bosonic fields, with the expectation that the fermions will come along for the ride.
 A notable exception is
 \cite{Nastase:1999cb,Nastase:1999kf} where a near complete reduction of $D=11$ supergravity on $S^4$ to maximal $D=7$ $SO(5)$ gauged
 supergravity was carried out.
 Also, in some cases \cite{Buchel:2006gb,Gauntlett:2006ai,gv} the fermions have been taken into consideration to the extent that
 one can show that the supersymmetry variations for bosonic configurations of the higher dimensional theory reduce to
 the expected supersymmetry variations for bosonic configurations in the lower dimensional theory. This allows one to
 obtain the useful result that any bosonic solution of the lower dimensional gauged supergravity that preserves
supersymmetry will uplift to a bosonic solution of $D=10$ or $D=11$ supergravity that also preserves supersymmetry.

It has recently been shown that the consistent KK reduction of $D=11$ supergravity on an arbitrary $SE_7$ space to minimal $D=4$ $N=2$
gauged supergravity that we mentioned above can be
generalised \cite{Gauntlett:2009zw}. At the level of bosonic fields it can be shown that in addition to the massless graviton supermultiplet one can also
include the massive supermultiplet that contains the breathing mode. The resulting consistent KK reduction gives a
$D=4$ $N=2$ gauged supergravity
coupled to a vector multiplet and a tensor multiplet. This matter content and the supersymmetry
can be understood in the following way.
First recall that one can consistently
KK reduce type IIA supergravity on an arbitrary Calabi-Yau three-fold to obtain a universal $D=4$ $N=2$ (ungauged) supergravity coupled
to universal tensor multiplet and a universal hypermultiplet. The details of this reduction utilise the fact that the Calabi-Yau three-fold
has an $SU(3)$ structure, specified by the K\"ahler form and the $(3,0)$ form, both of which are closed.
Returning to the reduction of $D=11$ supergravity on the $SE_7$ we next recall that it also has a globally defined $SU(3)$ structure
which implies that, locally, the $SE_7$ space can be viewed as
a $U(1)$ fibration over a six-dimensional K\"ahler-Einstein space $KE_6$. Thus, after first reducing on the $U(1)$,
the reduction has the structure of a type IIA reduction
on an $SU(3)$ manifold \cite{Grana}, the $KE_6$ space. Thus one expects the same field content and the same off-shell supersymmetry as in the universal sector of the
reduction of type IIA on the $CY_3$ space,
but the twisting of the $U(1)$ fibration, the fact that the $(3,0)$ form on the $KE_6$ is not closed and the presence of the background
four-form flux lead to a gauging of the $D=4$ $N=2$ supergravity theory. See \cite{O'Colgain:2009yd} for a related reduction of $D=11$ supergravity in a non-supersymmetric context.

In this paper we will show that there is an analogous generalisation of the KK reduction of type IIB supergravity on $SE_5$.
We will show that the consistent KK reduction to minimal $D=5$ $N=2$ gauged supergravity of \cite{Buchel:2006gb}
can also be extended to include
the massive supermultiplet containing the breathing mode. We will show that the reduction is
consistent with $D=5$ $N=4$ gauged supergravity
coupled to two vector multiplets with a gauging as described in \cite{Dall'Agata:2001vb}\cite{Schon:2006kz}.
To understand this matter content, and the origin of the increased supersymmetry, we now view,
locally, the $SE_5$  space as a $U(1)$ fibration over a four-dimensional K\"ahler-Einstein space, $KE_4$. The previous discussion suggests that
we should expect a gauged supergravity with the same field content and supersymmetry as that arising
in the universal sector of the reduction of type IIB supergravity on $HK_4\times S^1$,
where $HK_4$ is an arbitrary four-dimensional hyper-K\"ahler space (not necessarily compact). In fact this reduction has not yet been
analysed\footnote{For related work, see \cite{Lu:1998xt}.}, so in this paper we will show that  there is a consistent KK reduction of type IIB on $HK_4\times S^1$ to a universal sector that is consistent
with $N=4$ (ungauged) supergravity coupled to two vector multiplets.
For the reduction on $SE_5$ the twisting of the $U(1)$ fibration, the fact that the $(2,0)$ form on the $KE_4$ is not
closed and the presence of the background five-form flux lead to a gauging of the $D=5$ $N=4$ supergravity theory. We show that the gauging is given by a $H_3\times U(1)\subset SO(5,2)$ subgroup of the duality symmetry group $SO(1,1)\times SO(5,2)$
of the ungauged theory, where $H_3$ is the three-dimensional Heisenberg group.

The $D=5$ $N=4$ gauged supergravity that we obtain from the reduction on an $SE_5$ space admits
a supersymmetric $AdS_5$ vacuum which uplifts to the supersymmetric $AdS_5\times SE_5$ solutions of type IIB.
An interesting feature is that this $AdS_5$ vacuum spontaneously partially\footnote{Note that a general analysis of
such partial supersymmetry breaking has recently been carried out
for general $N=2$ $D=4$ gauged supergravities in \cite{Louis:2009xd} .} breaks the $N=4$ supersymmetry down to $N=2$. There is also another $AdS_5$ vacuum that doesn't preserve any supersymmetry which
uplifts to the type IIB solutions found by Romans \cite{Romans:1984an} generalising those found in $D=11$ by
Pope and Warner \cite{Pope:1984bd}\cite{Pope:1984jj}.  Without supersymmetry, the stability of these type IIB
solutions should be investigated;
for the special case that the $SE_5$ space is the round $S^5$ it is expected that they are not stable
\cite{hpw}.

The plan of the rest of the paper is as follows. The reduction of type IIB supergravity on $HK_4\times S^1$ is analysed in section 2
and the reduction on $SE_5$ is analysed in section 3. We have included some details of our calculations, which are rather long,
in an appendix. Section 4 concludes with some brief final comments concerning how our results might be generalised for the special case of $S^5$.
We also briefly comment on the consistent KK reduction of $D=11$ supergravity on
tri-Sasaki manifolds and argue that it will lead to an $N=4$ gauged supergravity in $D=4$ with an $AdS_4$ vacuum that spontaneously breaks
$N=4$ to $N=3$.

\subsection*{Note Added}
When we were writing this work up we became aware of \cite{Cassani:2010uw} with which
there is considerable overlap. On the same day that this
paper was posted to the arXive, \cite{Liu:2010sa, Skenderis:2010vz} appeared with which there is also overlap.

\vspace{1cm}

\section{Type IIB reduced on $HK_4\times S^1$ }
Our starting point is the class of $\bbR^{1,4}\times HK_4\times S^1$ solutions of type IIB supergravity.
Recall that the bosonic fields of type IIB supergravity
\cite{Schwarz:1983qr}\cite{Howe:1983sr}
consist of the metric, the dilaton $\Phi$ and the NS three-form field strength $H_{(3)}$, and the
RR form field-strengths $F_{(1)}=dC_{(0)}$, $F_{(3)}$, and $F_{(5)}$. The equations of motion
and Bianchi identities are given in appendix \ref{appa}.
The $\bbR^{1,4}\times HK_4\times S^1$ solution is given by
\bea\label{hksol}
ds^2_{10}&=&ds^2(\bbR^{1,4})+ds^2(\textrm{$HK_4$})+\eta\otimes \eta
\eea
with $F_{(5)}=F_{(3)}=H_{(3)}=0$ and has constant dilaton and constant axion ($F_{(1)}=0$).
The hyper-K\"ahler space $HK_4$ has a K\"ahler two-form $J$ and a $(2,0)$ form
$\Omega$  that satisfy algebraic conditions that are given in appendix \ref{appb}.
They are both closed as is the one-form $\eta$ on the $S^1$ factor:
\bea
dJ=d\Omega=d\eta=0
\eea
This solution generically preserves $N=4$ supersymmetry. Note that if $HK_4$ is compact then it
is either $K_3$ or $T^4$ and in the latter case all $N=8$ supersymmetry is preserved.

 \subsection{The consistent Kaluza-Klein reduction on $HK_4\times S^1$}

Our KK ansatz for the metric of type IIB supergravity
is given by
\be\label{KKmethk}
ds^2_{10}=e^{-\frac{2}{3}(4U+V)}ds^2_{(E)} +e^{2U}ds^2(\textrm{$HK_4$})+e^{2V}(\eta+A_1)\otimes(\eta +A_1) \ee
where $ds^2_{(E)}$ is an arbitrary metric on an external five-dimensional space-time
(it will turn out to be in the Einstein frame and hence the subscript $E$), $U$ and $V$ are scalar fields
and $A_1$ is a one-form defined on the external five-dimensional space. Following \cite{Gauntlett:2009zw}, the ansatz for the form field strengths is constructed using the two-forms $J,\Omega$ and $\eta$, and
is given by
{\setlength\arraycolsep{1pt}
\begin{eqnarray} \label{KKformshk}
F_{(5)} &=&
%4 e^{-\frac{8}{3}(4U+V)} \textrm{vol}^E_5 +
e^{-\frac{4}{3}(U+V)} * K_2 \wedge J+ K_1 \wedge J \wedge J\nn
&&+ \left[
%2 J \wedge J
-2e^{-8U} *K_1   +K_2 \wedge J \right]\wedge (\eta+A_1)\nn
 && + \left[ e^{-\frac{4}{3}(U+V)} *L_2 \wedge \Omega + L_2 \wedge \Omega \wedge (\eta+A_1) +c.c. \right]
\nonumber \\[8pt]
F_{(3)} &=& G_3 +G_2 \wedge (\eta+A_1) +G_1 \wedge J
%+G_0 J \wedge \hat e^5
+\left( N_1 \wedge \Omega  +c.c. \right)
\nonumber \\[8pt]
 H_{(3)} &=& H_3 +H_2 \wedge (\eta+A_1) +H_1 \wedge J
%+H_0 J \wedge \hat e^5
+\left( M_1 \wedge \Omega  +c.c. \right)
\nonumber \\[8pt]
C_{(0)} &=& a
\nonumber \\[8pt]
\Phi &=& \phi
\end{eqnarray}
Here, $*$ is the Hodge dual corresponding to the five-dimensional metric $ds^2_{(E)}$ in (\ref{KKmethk}), with volume form $\textrm{vol}^{(E)}_5$;
$a$, $\phi$,
are real scalars, $G_3$, $H_3$, $G_2$, $H_2$, $G_1$, $H_1$, $K_2$, $K_1$  real forms, and  $L_2$, $M_1$, $N_1$, complex forms, all of them defined on the external five-dimensional spacetime; and $c.c.$ denotes complex conjugate. Note that we have ensured that the five-form $F_{(5)}$ is self-dual with respect to the metric (\ref{KKmethk}).
Also note that we can add the terms $(G_0 J +N_0\Omega)\wedge (\eta+A_1)$ to $F_{(3)}$ and
$(H_0 J +M_0\Omega)\wedge (\eta+A_1)$ to $H_{(3)}$, where
$G_0$,$H_0$ are real scalars and $N_0,M_0$ are complex scalars. However, an analysis of the type IIB supergravity equations
imply that they can be set to zero. Similarly we
have also set to zero a possible factor $e^Z$, where $Z$ is a scalar, that would multiply $\textrm{vol}^{(E)}_5$ and $J\wedge J\wedge (\eta+A_1)$ terms in $F_{(5)}$.

We now substitute into the equations of motion and Bianchi identities of type IIB supergravity that are given in appendix \ref{appa}.
The calculations are rather involved, so we will simply summarise the main results here, referring to appendix \ref{appb} for some details.
We find that the physical
degrees of freedom are 7 real scalars $U,V,\phi,a,b,c,h$; 2 complex scalars $\xi,\chi$;
4 real one-form potentials $A_1,B_1,C_1,E_1$; 1 complex one-form potential $D_1$ and  2 real two-form potentials $B_2,C_2$ with
\bea \label{fieldstr1}
H_3&=&dB_2 -B_1 \wedge F_2 \nn
H_2&=&dB_1\nn
H_1&=&db\nn
M_1&=&d\xi
\eea
where $F_2\equiv dA_1$,
\bea \label{fieldstr2}
G_3&=& dC_2 -C_1 \wedge F_2 -adB_2 + a B_1 \wedge  F_2\nn
G_2&=&dC_1-adB_1\nn
G_1&=& dc-adb \nn
N_1&=&d\chi-ad\xi
\eea
and
\begin{eqnarray} \label{fieldstr3}
K_2 &=&  dE_1 -cdB_1 +bdC_1  \nn
L_2 & =& dD_1 -\chi dB_1 + \xi  dC_1 \nn
K_1 &=& dh +\ft12 (bdc-cdb) + \xi^* d\chi+\xi d\chi^*-\chi d\xi^*-\chi^* d\xi
\end{eqnarray}

We also find that the equations of motion for all of the fields can be obtained
by varying a $D=5$ action with Lagrangian given by
\be\label{lagEinframeK3}
{\cal L}^{(E)}={\cal L}_{\textrm{kin}}^{(E)}+{\cal L}_{\textrm{top}}
\ee
where the kinetic term is given by
\begin{eqnarray} \label{kineticEinframeK3}
{\cal L}_{\textrm{kin}}^{(E)} &=& R^{(E)} \ \textrm{vol}_5^{(E)} -\tfrac{28}{3} dU \wedge *dU  -\tfrac{8}{3} dU \wedge *dV - \tfrac{4}{3} dV \wedge * dV  -\ft12 e^{2\phi} da \wedge * da \nonumber \\ &&  -\ft12 d\phi \wedge * d\phi
-4e^{-4U-\phi} M_1 \wedge * M^*_1 -4e^{-4U+\phi} N_1 \wedge * N^*_1
-2 e^{-8U} K_1 \wedge * K_1\nn
&&- e^{-4U-\phi} H_1 \wedge * H_1 - e^{-4U+\phi} G_1 \wedge * G_1
-\ft12 e^{\frac{8}{3}(U+V)} F_2\wedge * F_2 \nonumber  \\ &&
 - e^{-\frac{4}{3}(U+V)} K_2 \wedge *K_2 -4 e^{-\frac{4}{3}(U+V)} L_2 \wedge * L^*_2
 -\ft{1}{2} e^{\frac{4}{3}(2U-V)-\phi} H_2 \wedge * H_2
 \nonumber  \\ && -\ft{1}{2} e^{\frac{4}{3}(2U-V)+\phi} G_2 \wedge * G_2
 -\ft{1}{2} e^{\frac{4}{3}(4U+V)-\phi} H_3 \wedge * H_3 -\ft{1}{2} e^{\frac{4}{3}(4U+V)+\phi} G_3 \wedge * G_3
  \nonumber\\
 \end{eqnarray}
and the topological term is given by
{\setlength\arraycolsep{0pt}
\begin{eqnarray}
{\cal L}_{\textrm{top}} \; &=&
A_1 \wedge \Big[-K_2 \wedge K_2 -4L_2 \wedge L_2^* +2K_1 \wedge \big( C_1 \wedge dB_1 -B_1 \wedge dC_1 \big)  \nonumber \\
&& \qquad \quad -2 K_2 \wedge (B_1 \wedge dc- C_1 \wedge db) -\big[4 L_2^* \wedge \big( B_1 \wedge d\chi - C_1 \wedge d \xi \big) +c.c. \big]    \Big] \nonumber \\
&& -2dC_2 \wedge X_2 +2dB_2 \wedge Y_2
\end{eqnarray}
}
where
\begin{eqnarray} \label{X2Y2}
X_2 &=& \left(h +\tfrac12 bc +\xi^*\chi +\xi \chi^* \right) dB_1 -\left( \tfrac12 b^2 +2|\xi|^2  \right) dC_1 -bdE_1 -2 \xi^*dD_1-2 \xi dD_1^* \nonumber \\
Y_2 &=& \left(h -\tfrac12 bc -\xi^*\chi -\xi \chi^* \right) dC_1 +\left( \tfrac12 c^2 +2|\chi|^2  \right) dB_1 -cdE_1 -2 \chi^*dD_1-2 \chi dD_1^* \nonumber \\
\end{eqnarray}
To summarise, by explicit construction, we have shown that any solution of the equations of motion arising from this $D=5$
action can be uplifted on an arbitrary $HK_4\times S^1$ space
via \reef{KKmethk} and \reef{KKformshk} to obtain exact solutions of type IIB supergravity. In other words we have identified
the consistent KK reduction of type IIB supergravity on $HK_4\times S^1$ to a universal $D=5$ theory. In the next subsection we will argue
that this $D=5$ theory is consistent with the bosonic sector of $N=4$ supergravity coupled to two vector multiplets.

\subsection{$N=4$ ungauged supergravity} \label{secSusyUngauged}

We first recall some aspects of $N=4$ ungauged supergravity coupled to $n$ vector multiplets
\cite{Awada:1985ep}  (see also \cite{Dall'Agata:2001vb}\cite{Schon:2006kz} ).
%The bosonic content of the $N=4$ gravity multiplet consists of a metric plus six scalars transforming in the ${\bf 1}+{\bf 5}$ of the
%$USp(4)\cong SO(5)$ $R$-symmetry group. The bosonic content of each $N=4$ vector mutliplet consists
%of a vector plus five scalars transforming in the ${\bf 5}$. In $N=4$ supergravity coupled to $n$ vector
%multiplets
The global symmetry group of the theory is given by $SO(1,1)\times SO(5,n)$.
The bosonic content includes a metric and $6+n$ vector fields, ${\cal B}^0_1,{\cal B}^M_1$ with $M=1,\dots 5+n$ transforming in the $(-1,{\bf 1})$ and $(+1/2,{\bf 5+n})$ representation, where the first entry indicates the $SO(1,1)$ charge.
In addition there are $1+5n$ scalar fields which parametrise the coset
$SO(1,1)\times SO(5,n)/SO(5)\times SO(n)$. The scalar corresponding to the $SO(1,1)$ factor
is described by a real scalar field $\Sigma$ which is a singlet under $SO(5,n)$ and carries
$SO(1,1)$ charge $-1/2$. The remaining $5n$ scalars are described by a coset representative
${\cal V}$ of $SO(5,n)/SO(5)\times SO(n)$ with zero $SO(1,1)$ charge.
The bosonic action of $D=5$ $N=4$ supergravity can be written
\bea\label{ungaugedact}
{\cal L}^{N=4}&=&R\textrm{vol}_5^{(E)}-\Sigma^2M_{MN}{\cal H}^M_2\wedge *{\cal H}^{N}_2
-\Sigma^{-4}{\cal H}^0_2\wedge *{\cal H}^0_2\nn
&-&3\Sigma^{-2}d\Sigma \wedge *d\Sigma
+\tfrac{1}{8}\textrm{tr}(dM\wedge *dM)\nn
&+&{\sqrt 2}\eta_{MN}{\cal B}^0_1\wedge {\cal H}^{M}_2\wedge {\cal H}^N_2
\eea
where $M\equiv {\cal V}^T{\cal V}$,  ${\cal H}^0_2\equiv d{\cal B}^0_1$,
${\cal H}^M_2\equiv d{\cal B}^M_1$ and $\eta_{MN}$ is the invariant tensor of $SO(5,n)$.

We now show that our consistent truncation on $HK_4\times S^1$
is consistent with $N=4$ supersymmetry. This is, perhaps, to be anticipated.
The (integrable) $SU(2)$ structure, defined by the forms $J,\Omega,\eta$,
can be constructed from the type IIB Killing spinors. Moreover, all KK modes can be decomposed
into representations of this $SU(2)$ structure and will naturally fall into $N=4$ off-shell multiplets.
The KK ansatz that we have shown to be a consistent truncation, is in fact the most general
bosonic ansatz that can be
constructed from modes on $HK_4\times S^1$ that are singlets under $SU(2)$ and are also constant.
As a consequence we must have kept the bosonic fields of a complete $N=4$ multiplet.
Indeed, in addition to the metric, our $D=5$ reduced theory has eleven real scalar degrees of freedom,
six real vectors and two real two-forms. Since we can dualise the two-forms to vectors
we have the bosonic matter content of $N=4$ supergravity coupled to $n=2$ vector multiplets.

In order to make the supersymmetry structure of our theory manifest, we now
dualise the two-forms $B_2$ and $C_2$ into two vectors $C_1^\prime$ and $B_1^\prime$ by defining
$H_3^\prime =dB_2$ and $G_3^\prime =dC_2$, and adding the term
\begin{eqnarray}
{\cal L}^\prime = C_1^\prime \wedge dH_3^\prime +B_1^\prime \wedge
dG_3^\prime
\end{eqnarray}
to the Lagrangian ${\cal L}^{(E)}$ in (\ref{lagEinframeK3}). Integrating out $H_3^\prime$ and $G_3^\prime$, we find that $H_3$ and $G_3$ are now given by
\begin{eqnarray} \label{H3G3duals}
H_3 &=& -e^{-\frac{4}{3}(4U+V)+\phi}  *G_2^\prime \nonumber \\
G_3 &=& -e^{-\frac{4}{3}(4U+V)-\phi}* H_2^\prime
\end{eqnarray}
where we have defined
\bea
H_2^\prime &=&dB_1^\prime -2X_2\nn
G_2^\prime &=&dC_1'+2Y_2 +a dB_1^\prime -2aX_2
\eea
and  $X_2$, $Y_2$ are given in (\ref{X2Y2}).
Substituting $H_3, G_3$, as given in (\ref{H3G3duals}),  back into ${\cal L}^{(E)} + {\cal L}^\prime$ we obtain a dual Lagrangian
${\cal L}^{\textrm{dual}}$ which
contains eight vector fields. With a little further effort we can show that the topological Lagrangian
simplifies considerably and in particular all dependence on the scalar fields drops out.
Before writing this action we first introduce new scalar fields given by
\begin{eqnarray}
\Sigma = e^{-\tfrac{2}{3} (U+V)} \, , \quad \varphi_1 = \tfrac{1}{\sqrt 2} (\phi-4U) \, ,  \quad \varphi_2 = -\tfrac{1}{\sqrt 2} (\phi+4U) \; .
\end{eqnarray}
The dual Lagrangian can then be written as
\begin{eqnarray} \label{LdualK3}
{\cal L}^{\textrm{dual}} = R^{(E)}  \textrm{vol}_5^{(E)}
+{\cal L}_{\textrm{scalars}}+{\cal L}_{\textrm{vectors}}+{\cal L}^{\textrm{dual}}_{\textrm{top}}
\end{eqnarray}
where the scalar kinetic terms are given by
\begin{eqnarray} \label{scalarscan}
{\cal L}_{\textrm{scalars}} &=& -3  \Sigma^{-2}d\Sigma \wedge * d \Sigma  -\tfrac12  d\varphi_1 \wedge * d \varphi_1  -\tfrac12  d\varphi_2 \wedge * d \varphi_2 \nonumber \\
&& -\ft12 e^{\sqrt{2} (\varphi_1 - \varphi_2)} da \wedge * da
-2 e^{\sqrt{2} (\varphi_1 + \varphi_2)} K_1 \wedge * K_1 \nonumber \\
&& - e^{\sqrt{2} \varphi_1} G_1 \wedge * G_1
-4 e^{\sqrt{2} \varphi_1}  N_1 \wedge * N^*_1 \nonumber \\
&& - e^{\sqrt{2} \varphi_2} H_1 \wedge * H_1
-4 e^{\sqrt{2} \varphi_2} M_1 \wedge * M^*_1
      \; ,
\end{eqnarray}
the kinetic terms for the vectors are given by
\begin{eqnarray} \label{LdualkinK3vec2}
{\cal L}_{\textrm{vectors}} &=&  -\ft12 \Sigma^{-4} F_2\wedge * F_2 -
\Sigma^2 \Big[  K_2 \wedge *K_2 + 4  L_2 \wedge * L^*_2 +\ft{1}{2}
e^{\sqrt{2}\varphi_2} H_2^\prime \wedge * H_2^\prime \nonumber  \\ &&  +
\ft{1}{2} e^{\sqrt{2}\varphi_1} G_2^\prime \wedge * G_2^\prime
+\ft{1}{2} e^{-\sqrt{2}\varphi_1} H_2 \wedge * H_2  +\ft{1}{2}
e^{-\sqrt{2}\varphi_2} G_2 \wedge * G_2 \Big]
\end{eqnarray}
and the topological term is
\begin{eqnarray} \label{LdualtopK3simp}
{\cal L}^{\textrm{dual}}_{\textrm{top}} &=& -A_1 \wedge \Big[dE_1 \wedge dE_1 +4 dD_1 \wedge dD_1^* -dB_1 \wedge dC_1^\prime -dC_1 \wedge dB_1^\prime  \Big]
\end{eqnarray}

We can now identify with the degrees of freedom of $N=4$ supergravity.
For the scalars we see that $\Sigma$ corresponds to the $\mathbb{R} \sim SO(1,1)$ factor in the scalar manifold. The remaining dilatons, $\varphi_1$, $\varphi_2$, and the axions
$a$, $b$, $c$, $h$, $\xi$, $\chi$, therefore
parametrise the homogeneous space $SO(5,2)/(SO(5) \times SO(2))$. To make this manifest we find it convenient to resort to the solvable Lie algebra approach \cite{Andrianopoli:1996bq, Andrianopoli:1996zg}. According to this method, a  parametrisation of the supergravity scalar manifold $G/H$ can be obtained via the exponentiation of a suitable solvable subalgebra of the Lie algebra of $G$, including as many Cartan generators as dilatons, and as many positive root generators as axions that are contained in $G/H$. For $SO(5,2)/(SO(5) \times SO(2))$, the relevant ten-dimensional subalgebra of $so(7)$ is accordingly spanned by two Cartan generators $\textsf{H}^1$, $\textsf{H}^2$, and eight positive root generators $\textsf{T}^i$, $i=1 , \ldots , 8$, which, in the fundamental of $so(7)$, can be taken to be\footnote{These generators close into the (solvable) Lie algebra with commutators specified in (3.12) of \cite{Lu:1998xt}. With $N=3$ there, we identify the positive root generators as $\textsf{T}^1=E_2{}^3$, $\textsf{T}^2=V^{23}$, $\textsf{T}^3=U^2_1$,  $\textsf{T}^4=U^3_1$,  $\textsf{T}^5=U^2_2$, $\textsf{T}^6 = U^2_3$,  $\textsf{T}^7=U^3_2$, $\textsf{T}^8 = U^3_3$. Our explicit realisation (\ref{generators}) of these generators follows from (3.31) of  \cite{Lu:1998xt}.} \cite{Lu:1998xt}
\begin{equation} \label{generators}
\begin{tabular}{llll}
$\textsf{H}^1 = \sqrt{2} (E_{22}-E_{77})$, & $\textsf{T}^1 = E_{67} -E_{21}$,
            & $\textsf{T}^4 = E_{23}+E_{37}$,  & $\textsf{T}^7 = E_{24}+E_{47}$, \\
$\textsf{H}^2 = \sqrt{2}  (E_{11}-E_{66})$, & $\textsf{T}^2 = E_{17}-E_{26}$,
            & $\textsf{T}^5 = E_{14}+E_{46}$,  & $\textsf{T}^8 = E_{25}+E_{57}$, \\
 & $\textsf{T}^3 = E_{13}+E_{36}$,
            & $\textsf{T}^6 = E_{15}+E_{56}$, &
\end{tabular}
\end{equation}
where $E_{ij}$ denotes the $7 \times 7$ matrix with 1 in the $i$-th row and $j$-th column and 0 elsewhere.

We find the coset representative of $SO(5,2)/(SO(5) \times SO(2))$ to be given by
\begin{eqnarray} \label{coset}
{\cal V} &=& e^{\frac{1}{2} (\varphi_1 \textsf{H}^1 + \varphi_2 \textsf{H}^2)} e^{-a \textsf{T}^1} e^{ -(2h - bc -2\xi^* \chi -2\xi \chi^* ) \textsf{T}^2} e^{ b \sqrt{2} \textsf{T}^3}  e^{-c \sqrt{2} \textsf{T}^4} e^{2 \sqrt{2}\textrm{Re}(\xi) \textsf{T}^5} e^{2 \sqrt{2} \textrm{Im}(\xi)  \textsf{T}^6}  \nonumber \\ &&  \times e^{-2 \sqrt{2}\textrm{Re}(\chi) \textsf{T}^7} e^{-2 \sqrt{2}\textrm{Im}(\chi) \textsf{T}^8}  \; .
\end{eqnarray}
Note that in this basis we have ${\cal V}^T\eta{\cal V}=\eta$ with
$\eta=E_{33}+E_{44}+E_{55}-E_{16}-E_{61}-E_{27}-E_{72}$.
The Maurer-Cartan form $d {\cal V} {\cal V}^{-1} $ takes values in the solvable Lie algebra,
\begin{eqnarray}
d {\cal V}  {\cal V}^{-1} &=& \tfrac12 d \varphi_1 \textsf{H}^1 +\tfrac12 d \varphi_2 \textsf{H}^2 - e^{\frac{\sqrt{2}}{2}(\varphi_1 -\varphi_2)} da \textsf{T}^1 - 2 e^{\frac{\sqrt{2}}{2}(\varphi_1 +\varphi_2)} K_1 \textsf{T}^2 + \sqrt{2} e^{\frac{\sqrt{2}}{2}\varphi_2 } H_1  \textsf{T}^3 \nonumber \\
&&-  \sqrt{2} e^{\frac{\sqrt{2}}{2}\varphi_1 } G_1 \textsf{T}^4 + 2\sqrt{2} e^{\frac{\sqrt{2}}{2}\varphi_2 } \textrm{Re}(M_1) \textsf{T}^5 + 2\sqrt{2} e^{\frac{\sqrt{2}}{2}\varphi_2 } \textrm{Im}(M_1) \textsf{T}^6
\nonumber \\
&& - 2\sqrt{2} e^{\frac{\sqrt{2}}{2}\varphi_1} \textrm{Re}(N_1) \textsf{T}^7 - 2\sqrt{2} e^{\frac{\sqrt{2}}{2}\varphi_1} \textrm{Im}(N_1) \textsf{T}^8 \; ,
 \end{eqnarray}
with coefficients associated to the positive root generators in (\ref{generators}) corresponding to the axion one-form field strengths defined in  (\ref{fieldstr1})--(\ref{fieldstr3}). Note that the transgression terms in these one-forms arise as a consequence of the non-trivial commutation relations among the positive root generators.
Finally, we can construct the quadratic form
\begin{equation} \label{scalarmatrix}
M = {\cal V}^T{\cal V} \; ,
\end{equation}
to bring the scalar kinetic terms (\ref{scalarscan}) to the form
\begin{eqnarray} %\label{scalarscan}
{\cal L}_{\textrm{scalars}} &=& -3 \Sigma^{-2} d\Sigma  \wedge * d \Sigma  +\tfrac18  \textrm{tr} (d M^{-1}  \wedge * d M )
      \; .
\end{eqnarray}
exactly as in \reef{ungaugedact}.

For the vectors we identify
\bea \label{vectorrelabel}
{\cal B}^0_1&=& -\tfrac{1}{\sqrt 2}A_1\nn
{\cal B}_1^M &=& \{ \tfrac{1}{\sqrt 2}B_1'\; ,\; \tfrac{1}{\sqrt 2}C_1',\; ,\;E_1 , \;  2 \textrm{Re}(D_1)  , \;  2 \textrm{Im}(D_1)  , \;  \tfrac{1}{\sqrt 2}C_1 \; ,
\tfrac{1}{\sqrt 2}B_1   \;    \} \; ,
\eea
In particular, for our Chern-Simons term we then have
\begin{eqnarray} \label{LdualtopK3simpdiagfin}
{\cal L}^{\textrm{dual}}_{\textrm{top}} &=&  {\sqrt 2}\eta_{MN} {\cal B}^0_1 \wedge  {\cal H}_2^M \wedge {\cal H}_2^N \; .
\end{eqnarray}
and we have verified that
the kinetic terms for the vectors given in \reef{LdualkinK3vec2}
can be written as
\be
{\cal L}_{\textrm{vectors}} =-\Sigma^2M_{MN}{\cal H}^M_2\wedge *{\cal H}^{N}_2
-\Sigma^{-4}{\cal H}^0_2\wedge *{\cal H}^0_2
\ee
as in \reef{ungaugedact}.
This completes our demonstration that we indeed have the bosonic action of $N=4$ supergravity coupled to two vector multiplets.

\section{Type IIB reduced on $SE_5$}
We now turn our attention to reductions on $SE_5$ spaces. We begin by
recalling the class of $AdS_5\times SE_5$ solutions of type IIB supergravity given by
\bea\label{sesol}
ds^2_{10}&=&ds^2(AdS_5)+ds^2(\textrm{$SE_5$})\nn
%&=&ds^2(AdS_5)+ds^2(\textrm{$KE_4$})+\eta\otimes \eta\nn
F_{(5)}&=&4\textrm{vol}(SE_5)+4 \textrm{vol}(AdS_5)
\eea
with $F_{(3)}=H_{(3)}=0$ and constant dilaton and constant axion ($F_{(1)}=0$).
Generically these solutions preserve $N=2$ supersymmetry (i.e. dual to $N=1$ SCFTs in $d=4$).

Now any $SE_5$ space has a globally defined one-form $\eta$, that is dual to the Reeb Killing vector,
and so locally we can always write the metric on the $SE_5$ space as
\be
ds^2(SE_5)=ds^2(\textrm{$KE_4$})+\eta\otimes \eta
\ee
where $ds^2(KE_4)$ is a local K\"ahler-Einstein metric with positive curvature, normalised
so that the Ricci tensor is six times the metric. On $SE_5$ there is also a globally defined
two form $J$ and a $(2,0)$ form $\Omega$ that locally define the K\"ahler and complex structures on
$ds^2(KE_4)$ respectively. Note that they satisfy the {\it same} algebraic conditions as
those associated with the $HK_4\times S^1$ solution \reef{hksol} and are given in appendix B.
By contrast, however, they are no longer closed and
instead satisfy
\bea\label{diffconds}
d\eta&=&2J\nn
d\Omega&=&3i\eta\wedge \Omega
\eea

The fact that $HK_4\times S^1$ and $SE_5$ have globally defined $SU(2)$ structures specified by the forms $J,\Omega,\eta$ implies that the universal KK reduction on these spaces are very similar, as
we shall see. For the $SE_5$ case the
conditions \reef{diffconds} as well as the background five-form flux appearing in \reef{sesol} will lead to a gauging
of the $N=4$ supergravity coupled to two vector multiplets that we saw for the $HK_4\times S^1$ case
in the last section.

\subsection{The consistent Kaluza-Klein reduction on $SE_5$}
Our KK ansatz for the metric of type IIB supergravity is given by
\be\label{KKmet}
ds^2_{10}=e^{-\frac{2}{3}(4U+V)}ds^2_{(E)}+e^{2U}ds^2(\textrm{$KE_4$})+e^{2V}(\eta+A_1)\otimes(\eta +A_1) \ee
where again $ds^2_{(E)}$ is an arbitrary metric on an external five-dimensional space-time,
$U$ and $V$ are scalar fields and $A_1$ is a one-form defined on the external five-dimensional space.
The ansatz for the form field strengths is given by
{\setlength\arraycolsep{1pt}
\begin{eqnarray} \label{KKforms}
F_{(5)} &=& 4 e^{-\frac{8}{3}(4U+V)+Z} \textrm{vol}^{(E)}_5 +e^{-\frac{4}{3}(U+V)} * K_2 \wedge J+ K_1 \wedge J \wedge J\nn
&&+ \left[2e^Z J \wedge J  -2e^{-8U} *K_1   +K_2 \wedge J \right]\wedge (\eta+A_1)\nn
 && + \left[ e^{-\frac{4}{3}(U+V)} *L_2 \wedge \Omega + L_2 \wedge \Omega \wedge (\eta+A_1) +c.c. \right]
\nonumber \\[8pt]
F_{(3)} &=& G_3 +G_2 \wedge (\eta+A_1) +G_1 \wedge J
%+G_0 J \wedge \hat e^5
+\left[ N_1 \wedge \Omega +N_0 \Omega \wedge (\eta+A_1) +c.c. \right]
\nonumber \\[8pt]
 H_{(3)} &=& H_3 +H_2 \wedge (\eta+A_1) +H_1 \wedge J
%+H_0 J \wedge \hat e^5
+\left[ M_1 \wedge \Omega +M_0 \Omega \wedge (\eta+A_1) +c.c. \right]
\nonumber \\[8pt]
C_{(0)} &=& a
\nonumber \\[8pt]
\Phi &=& \phi
\end{eqnarray}
}Here, $\textrm{vol}^{(E)}_5$ and $*$ are the volume form and Hodge dual corresponding to the five-dimensional metric
$ds^2_{(E)}$ in (\ref{KKmet}), $Z$, $a$, $\phi$,
are real scalars, $M_0$, $N_0$ complex scalars,  $G_3$, $H_3$, $G_2$, $H_2$, $G_1$, $H_1$, $K_2$, $K_1$  real forms, and  $L_2$, $M_1$, $N_1$, complex forms, all of them defined on the external five-dimensional spacetime. We have also ensured the self duality of the five-form $F_{(5)}$
with respect to the metric (\ref{KKmet}). Note that we can also add the terms $G_0 J \wedge (\eta+A_1)$ to $F_{(3)}$ and $H_0 J \wedge (\eta+A_1)$ to $H_{(3)}$, where
$G_0$ and  $H_0$ are real scalars, but the type IIB equations imply that $G_0=H_0=0$. This is as in the $HK_4\times S^1$ case, but, by contrast, note that we now must include the scalar
fields $M_0$, $N_0$ and $Z$.

We now substitute into the equations of motion and Bianchi identities of type IIB supergravity that we have given in appendix \ref{appa}.
The calculations are rather involved, so we will simply summarise the main results here, referring to appendix \ref{appb} for some details.
We find that the physical
degrees of freedom are 7 real scalars $U,V,\phi,a,b,c,h$ and 2 complex scalars $\xi,\chi$;
4 one-form potentials $A_1,B_1,C_1,E_1$; 2 two-form potentials $B_2,C_2$ plus the complex two-form $L_2$. This is exactly the same field content that arose in the reduction on $HK_4\times S^1$.
In particular, the extra scalars $M_0$, $N_0$ and $Z$ that we introduced in \reef{KKforms} are
not independent degrees of freedom as we shall see in detail below. Furthermore, the field
$L_2$ should also be regarded as a potential, satisfying a first-order, self-duality-type equation of motion
(see the second equation in (\ref{eqsfromF5})).

In more detail we find that
\bea\label{scone}
H_3&=&
%dB_2+\ft12 H_1\wedge F_2=
dB_2+\tfrac12(db-2B_1) \wedge F_2 \nn
H_2&=&dB_1\nn
H_1&=&db-2B_1\nn
M_0&=&3i\xi\nn
M_1&=&D\xi
\eea
and
\bea\label{sctwo}
G_3&=&
%dC_2 +\tfrac12 G_1 \wedge F_2 -a (H_3 -\tfrac12 H_1 \wedge F_2) =
dC_2 -adB_2 + \tfrac12 (dc-2C_1-adb+2aB_1) \wedge F_2\nn
G_2&=&
%dC_1-aH_2=
dC_1-adB_1\nn
G_1&=&
%dc-2C_1-aH_1=
dc-2C_1-adb+2aB_1\nn
N_0&=&3i(\chi-a\xi)\nn
N_1&=&D\chi-aD\xi
\eea
where $F_2=dA_1$, $D\xi\equiv d\xi-3iA_1\xi$ and $D\chi\equiv d\chi-3iA_1\chi$. This gauging can be traced back to the fact that
these scalar fields enter the KK ansatz \reef{KKforms} with $\Omega$ and that $d\Omega=3i\eta\wedge\Omega$. In addition, after fixing
an integration constant, we find that the scalar field $Z$ is fixed by the scalars $\chi,\xi$:
\begin{equation}
e^Z = 1 +3i(\xi^*\chi-\xi\chi^*)
\end{equation}
and that
\begin{eqnarray}\label{scthree}
K_2
%&=& dE_1 +\tfrac12 H_1 \wedge G_1 \nonumber \\
&=&dE_1+\tfrac12(db-2B_1)\wedge(dc-2C_1)\nn
K_1
%&=& dh -2E_1 -2\lambda A_1 +\tfrac{i}{3} [M_0^*N_1-M_0N_1^*+N_0M_1^*-N_0^*M_1]\nn
&=&dh -2E_1 -2
%\lambda
A_1 +\xi^* D\chi+\xi D\chi^*-\chi D\xi^*-\chi^* D\xi
\end{eqnarray}

We also find that the equations of motion for all of the fields can be obtained
by varying a $D=5$ action with Lagrangian given by
\begin{equation} \label{lagrangianse}
{\cal L} = {\cal L}_{\textrm{kin}} + {\cal L}_{\textrm{pot}} + {\cal L}_{\textrm{top}}
\end{equation}
where the kinetic and scalar potential terms are given by
\begin{eqnarray} \label{kineticEinframeSE}
{\cal L}_{\textrm{kin}}^{(E)} &=& R^{(E)} \ \textrm{vol}_5^{(E)} -\tfrac{28}{3} dU \wedge *dU  -\tfrac{8}{3} dU \wedge *dV - \tfrac{4}{3} dV \wedge * dV -\ft12 e^{2\phi} da \wedge * da \nonumber \\ && -\ft12 d\phi \wedge * d\phi
-4e^{-4U-\phi} M_1 \wedge * M^*_1 -4e^{-4U+\phi} N_1 \wedge * N^*_1
-2 e^{-8U} K_1 \wedge * K_1\nn
&&- e^{-4U-\phi} H_1 \wedge * H_1 - e^{-4U+\phi} G_1 \wedge * G_1
-\ft12 e^{\frac{8}{3}(U+V)} F_2\wedge * F_2 \nonumber  \\ &&
 - e^{-\frac{4}{3}(U+V)} K_2 \wedge *K_2 -4 e^{-\frac{4}{3}(U+V)} L_2 \wedge * L^*_2
 -\ft{1}{2} e^{\frac{4}{3}(2U-V)-\phi} H_2 \wedge * H_2
 \nonumber  \\ && -\ft{1}{2} e^{\frac{4}{3}(2U-V)+\phi} G_2 \wedge * G_2
 -\ft{1}{2} e^{\frac{4}{3}(4U+V)-\phi} H_3 \wedge * H_3 -\ft{1}{2} e^{\frac{4}{3}(4U+V)+\phi} G_3 \wedge * G_3
  \nonumber\\
 \end{eqnarray}
and
\begin{eqnarray} \label{potentialEinframe}
{\cal L}_{\textrm{pot}}^{(E)} &=& \Big[ 24e^{-\frac{2}{3}(7U+V)} -4e^{\frac{4}{3}(-5U+V)} -8e^{-\frac{8}{3}(4U+V)} \left[
%\lambda
1+\ft{i}{3}(M_0^*N_0 -M_0N_0^*) \right]^2 \nonumber \\ && \; -4e^{-\frac{4}{3}(5U+2V)-\phi}|M_0|^2 -4e^{-\frac{4}{3}(5U+2V)+\phi}|N_0|^2 \Big] \textrm{vol}_5^{(E)} \; ,
  \end{eqnarray}
while the topological terms are given by the intimidating expression
\begin{eqnarray}\label{ltopmmm}
{\cal L}_{\textrm{top}}&=& -A_1\wedge K_2\wedge K_2 -(dh -2E_1 -2
%\lambda
A_1) \wedge [dB_2 \wedge (dc-2C_1) + (db-2B_1) \wedge dC_2] \nonumber \\ &&
+A_1 \wedge (dh -2E_1) \wedge [(db-2B_1) \wedge dC_1 -dB_1 \wedge (dc-2C_1)] \nonumber \\ && +2A_1 \wedge dE_1 \wedge (db-2B_1) \wedge (dc-2C_1) \nonumber
\\ && +
%\lambda
A_1 \wedge (db-2B_1) \wedge (dc-2C_1) \wedge F_2 \nonumber \\ && +\tfrac{i}{3} A_1 \wedge (M_0^* N_1 -M_0 N_1^* -N_0 M_1^* +N_0^*M_1)
\wedge (H_1\wedge G_2 +G_1 \wedge H_2) \nonumber \\ && +\tfrac{2i}{3} A_1 \wedge (N_0N_1^* -N_0^*N_1) \wedge H_1 \wedge H_2 +\tfrac{2i}{3} A_1 \wedge (M_0M_1^* -M_0^*M_1)
\wedge G_1 \wedge G_2 \nonumber \\ && +\Big[ \tfrac{4i}{3} \big(-\tfrac{1}{2} L_2^* \wedge DL_2 + N_0 L_2^* \wedge H_3 -M_0L_2^* \wedge G_3 -L_2^*
\wedge N_1 \wedge H_2 +L_2^*\wedge M_1 \wedge G_2 \big) +c.c. \Big] \nonumber \\ && -4
%\lambda
C_2 \wedge dB_2 -\tfrac{4i}{3} C_2 \wedge G_2
\wedge (M_0M_1^*-M_0^*M_1) \nonumber \\ && -\tfrac{2i}{3} C_2 \wedge H_2 \wedge (M_0^*N_1 -M_0N_1^* -N_0M_1^*+N_0^*M_1) \nonumber \\ &&
-\tfrac{2i}{3} B_2 \wedge (G_2 -aH_2) \wedge (M_0^*N_1 -M_0N_1^* -N_0M_1^*+N_0^*M_1) \nonumber \\ && -\tfrac{4i}{3} B_2 \wedge \big[ H_2
\wedge (N_0N_1^* -N_0^*N_1) -aG_2 \wedge (M_0M_1^*-M_0^*M_1) ]  \nonumber \\ &&
+\tfrac{4}{9} A_1 \wedge (M_0G_2 -N_0H_2) \wedge (M_0^*G_2 -N_0^*H_2) \nonumber \\ && + \big[\tfrac{2}{9} (M_1 \wedge G_2 -N_1 \wedge H_2)
\wedge (M_0^*G_2 -N_0^*H_2) + c.c. \big] \; .
\end{eqnarray}

\subsection{$N=4$ gauged supergravity}
Our KK reduction of type IIB supergravity was based on the most general ansatz using
the $SU(2)$ structure $(J,\Omega,\eta)$ on the $SE_5$ space. For reasons similar to that
discussed in section \ref{secSusyUngauged} for the universal KK reduction on $HK_4\times S^1$ we expect
to obtain a supersymmetric theory. Indeed we have already noted that the field content
of our KK reduction on $SE_5$ is identical to that of the universal KK reduction on $HK_4\times S^1$ and hence to that of $N=4$ supergravity coupled to two vector multiplets. A key difference with the
$HK_4\times S^1$ case is that the $SU(2)$ structure forms are no longer closed, see \reef{diffconds}.
This difference can be viewed as the addition of ``geometric fluxes". Another difference is
the extra five-form flux that we have in \reef{KKforms} compared with \reef{KKformshk}.
On rather general grounds \cite{Grana} it is expected that these flux contributions lead to a
gauging of the $N=4$ supergravity theory.

A general description of $N=4$ gauged supergravity coupled to vector multiplets
is presented in \cite{Schon:2006kz}, which uses the embedding tensor formalism
of \cite{deWit:2002vt} (for a review see  \cite{Samtleben:2008pe}). The gauging is described by
promoting a subgroup $G_0\subset G$ of the global non-abelian duality symmetry group of the ungauged theory, $G$
(in our case $SO(1,1)\times SO(5,2)$), to a local symmetry. This requires that the ordinary derivatives get
replaced by covariant derivatives via
\be\label{samt}
d\to d-g{\cal B}_1^{\cal M} X_{\cal M}\equiv d-g{\cal B}_1^{\cal M} \Theta_{\cal M}{}^\alpha t_\alpha
\ee
where $g$ is the gauge coupling constant, ${\cal B}_1^{\cal M}\equiv({\cal B}_1^0,{\cal B}_1^M )$ are the vector gauge fields,
$t_\alpha$ are the generators of $G$
and the explicit embedding of $G_0$ in $G$ is given by an embedding tensor $\Theta_{\cal M}{}^\alpha$. In addition to the vector gauge fields it
is necessary to also include two-form gauge fields $B^{\cal M}_2$ as off-shell degrees of freedom.
As usual, the gauging leads to a scalar potential that is determined by the embedding tensor.

We now return to our $D=5$ theory \reef{lagrangianse}-\reef{ltopmmm}
obtained from KK reduction on $SE_5$.
By again introducing
\begin{eqnarray}
\Sigma = e^{-\tfrac{2}{3} (U+V)} \, , \quad \varphi_1 = \tfrac{\sqrt{2}}{2} (\phi-4U) \, ,  \quad \varphi_2 = -\tfrac{\sqrt{2}}{2} (\phi+4U) \; .
\end{eqnarray}
we find that
the Lagrangian can  be written as
\begin{eqnarray} \label{LdualSE}
{\cal L} ^{(E)}= R^{(E)}  \textrm{vol}_5^{(E)}
+{\cal L}_{\textrm{scalars}}+{\cal L}_{\textrm{vectors/2-forms}}+{\cal L}^{(E)}_{\textrm{pot}}+{\cal L}_{\textrm{top}}
\end{eqnarray}
where the scalar kinetic terms are given by
\begin{eqnarray} \label{scalarscanSE}
{\cal L}_{\textrm{scalars}} &=& -3  \Sigma^{-2}d\Sigma \wedge * d \Sigma  -\tfrac12  d\varphi_1 \wedge * d \varphi_1  -\tfrac12  d\varphi_2 \wedge * d \varphi_2 \nonumber \\
&& -\ft12 e^{\sqrt{2} (\varphi_1 - \varphi_2)} da \wedge * da
-2 e^{\sqrt{2} (\varphi_1 + \varphi_2)} K_1 \wedge * K_1 \nonumber \\
&& - e^{\sqrt{2} \varphi_1} G_1 \wedge * G_1
-4 e^{\sqrt{2} \varphi_1}  N_1 \wedge * N^*_1 \nonumber \\
&& - e^{\sqrt{2} \varphi_2} H_1 \wedge * H_1
-4 e^{\sqrt{2} \varphi_2} M_1 \wedge * M^*_1
      \; .
\end{eqnarray}
the kinetic terms for the vectors and two-forms are given by
\begin{eqnarray} \label{LdualkinSEvec2}
{\cal L}_{\textrm{vectors/2-forms}} &=&  -\ft12 \Sigma^{-4} F_2\wedge * F_2 -
\Sigma^2 \Big[  K_2 \wedge *K_2 + 4  L_2 \wedge * L^*_2 +\ft{1}{2}
e^{\sqrt{2}\varphi_2} H_2^\prime \wedge * H_2^\prime \nonumber  \\ &&  +
\ft{1}{2} e^{\sqrt{2}\varphi_1} G_2^\prime \wedge * G_2^\prime
+\ft{1}{2} e^{-\sqrt{2}\varphi_1} H_2 \wedge * H_2  +\ft{1}{2}
e^{-\sqrt{2}\varphi_2} G_2 \wedge * G_2 \Big]
\end{eqnarray}
where $H_2^\prime$, $G_2^\prime$ are obtained from the two-form potentials $B_2$, $C_2$ by a dualisation, and the potential (\ref{potentialEinframe}) can be rewritten as
\begin{eqnarray}
{\cal L}_{\textrm{pot}}^{(E)} &=& \Big[ 24 \Sigma e^{\sqrt{2}(\varphi_1+\varphi_2)} -4\Sigma^{-2} e^{\sqrt{2}(\varphi_1+\varphi_2)}  \nonumber  \\
&& -4\Sigma^4 \big[ 2 e^{\sqrt{2}(\varphi_1+\varphi_2)} \big(1  +3i(\xi^*\chi-\xi\chi^*) \big)^2 +9e^{\sqrt{2} \varphi_1} |\chi-a\xi|^2 +9e^{\sqrt{2} \varphi_2} |\xi|^2 \big] \Big] \textrm{vol}_5^{(E)} \; .  \nonumber  \\
  \end{eqnarray}

We observe that the general structure of all these terms is consistent with the general form of the corresponding terms
in the $N=4$ gauged supergravity action given in \cite{Schon:2006kz}. Furthermore,
noting that the covariant derivative acting on $\Sigma$ in \reef{scalarscanSE} is simply the ordinary covariant
derivative, and comparing with \reef{samt} we immediately conclude that the gauging does not lie
within the $SO(1,1)$ factor
but just within the $SO(5,2)$ factor (and hence is in the class considered by
\cite{Dall'Agata:2001vb}).

By analysing the way in which the vector fields are entering
the scalar derivatives in \reef{scone}, \reef{sctwo} and \reef{scthree} and comparing with \reef{samt}
it is straightforward to deduce the precise gauged subgroup of $SO(5,2)$. Setting $g=1$ in (\ref{samt}), we find that
\begin{eqnarray}
{\cal B}^0_1 X_0 + {\cal B}^M_1 X_M & =& \sqrt{2} {\cal B}^0_1 (3\textsf{R} +4 \textsf{S}^2) -4 {\cal B}^3_1 \ \textsf{S}^2 + 4{\cal B}^7_1 \ \textsf{S}^3 -4{\cal B}^6_1 \ \textsf{S}^4  \nonumber \\
&=& -A_1 (3\textsf{R} +4 \textsf{S}^2) -4 E_1 \ \textsf{S}^2 + 2\sqrt{2}B_1 \ \textsf{S}^3 -2\sqrt{2} C_1 \ \textsf{S}^4 \; ,
\end{eqnarray}
where $\textsf{S}^i \equiv [\textsf{T}^i]^T$, $i=1, \ldots, 8$, and $\textsf{R}\equiv E_{45}-E_{54}$ are generators of $SO(5,2)$ supplementing those in
\reef{generators}.
The only non-vanishing commutators among the generators of the gauge algebra is $[\textsf{S}^3,\textsf{S}^4]=-\textsf{S}^2$ and hence
we see that our $D=5$ theory corresponds to a gauging of an $H_3\times U(1)$ subgroup of $SO(5,2)$, where $H_3$ is
the three-dimensional Heisenberg group.

It would be satisfying to see that the rest of our Lagrangian
is in accord with \cite{Dall'Agata:2001vb}\cite{Schon:2006kz}, especially our Chern-Simons terms \reef{ltopmmm}, but we leave that to future work.

\subsection{$AdS_5$ vacua}
By analysing the scalar potential \reef{potentialEinframe} appearing in our $D=5$ $N=4$ gauged supergravity theory we find that there are
both supersymmetric and non-supersymmetric $AdS_5$ vacua.
We first discuss the former and then the latter.

\subsubsection{The supersymmetric $AdS_5$ vacuum}
Setting $U=V=\xi=\chi=0$ and allowing for arbitrary constant axion $a$ and dilaton $\phi$ we obtain
an $AdS_5$ vacuum solution with unit radius (and all other fields trivial). This solution uplifts to the class of supersymmetric $AdS_5\times SE_5$ solutions of
type IIB given in \reef{sesol}. As a ten-dimensional solution this preserves eight supercharges and hence as a five dimensional
solution it spontaneously partially breaks the $N=4$ supersymmetry to $N=2$.

We can determine the masses of the different fields in this vacuum.
For the scalars $\phi$, $a$, $U$, $V$, $\xi$, $\chi$, we employ
\begin{eqnarray}
&& U = \ft12 u  + \ft34 v \nonumber \\
&& V = -2u + \ft34 v \; .
\end{eqnarray}
and
\begin{eqnarray} \label{MassesCharged}
&& \xi = \tilde \xi + i
%\lambda
\tilde  \chi \nonumber \\
&& \chi = i
%\lambda
\tilde  \xi + \tilde  \chi \; .
\end{eqnarray}
to obtain the masses:
\begin{equation}
m^2_\phi =0 \; , \; \; m^2_a =0 \; , \; \; m^2_u =12 \; , \; \; m^2_v =32 \; , \; \; m^2_{\tilde \xi} =-3 \; , \; \; m^2_{\tilde \chi} =21 \; , \quad
\end{equation}
Observe that $v$ is a breathing mode which controls the overall volume of the $SE_5$ space in \reef{KKmet}, while
$u$ is a volume preserving squashing mode.

Turning now to the contributions from the vectors $A_1$, $E_1$, $B_1$, $C_1$ and the scalars $h$, $b$, $c$,
we find that the transformation
\begin{eqnarray} \label{MassesVectors}
A_1 = \tilde A_1 +2
%\lambda
\tilde E_1 \nonumber \\
E_1 = -
%\lambda
\tilde A_1 + \tilde E_1
\end{eqnarray}
leads to the following terms in the Lagrangian
\begin{eqnarray}
{\cal L}_2=&&-\tfrac{3}{2} d\tilde A_1 \wedge *d\tilde A_1 -3  d \tilde E_1 \wedge *  d \tilde E_1  -2(dh -6 \tilde E_1) \wedge * (dh -6 \tilde E_1)  \nonumber \\
&& -\tfrac{1}{2} dB_1 \wedge *dB_1 -\tfrac{1}{2} dC_1 \wedge *dC_1  -(db-2B_1) \wedge * (db-2B_1)  \nonumber \\
&& -(dc-2C_1) \wedge * (dc-2C_1) \; .
\end{eqnarray}
We now see that $\tilde A_1$, $\tilde E_1$, $B_1$ and $C_1$ are massive vectors with masses given by
\begin{equation}
m^2_{\tilde A_1} =0 \; , \qquad m^2_{\tilde E_1} =24 \; , \qquad m^2_{B_1} = m^2_{C_1} =8 \; ,
\end{equation}
and that the scalars $h$, $b$ and $c$ are just the associated St\"uckelberg fields.

Next consider the two-forms $B_2$ and $C_2$, whose relevant contributions are given by
\begin{equation} \label{quad3}
{\cal L}_3 = -\tfrac12 dB_2 \wedge *dB_2 -\tfrac12 dC_2 \wedge *dC_2 -4
%\lambda
C_2 \wedge dB_2 \; ,
\end{equation}
they combine to describe a massive two-form with mass
\begin{equation}
m^2_{C_2} = 16 \;
\end{equation}
(see {\it e.g.} \cite{Minahan:1989vc}).
Finally, from the contribution
\begin{equation}
{\cal L}_4 = -\tfrac{2i}{3} L_2^* \wedge dL_2 +\tfrac{2i}{3} L_2 \wedge dL_2^* -4L_2 \wedge *L_2^*
\end{equation}
we see that $L_2$ is a complex two-form satisfying a self-duality equation
\cite{Townsend:1983xs} so that it has the same degrees of freedom as a massive real two-form\footnote{To see this
simply write $L_2$ as a real and an imaginary two form an observe that either one can be considered a Lagrange multiplier and
eliminated.} with
\begin{equation}
m^2_{L_2} =9 \; .
\end{equation}

To conclude, we quote here the scaling dimensions of the operators dual to the supergravity fields. Using the expressions
\begin{equation}
\Delta = 2 \pm \ft12 \sqrt{(4-2p)^2 +4m^2} \; ,
\end{equation}
for four-dimensional operators dual to supergravity $p$-forms (subject to second-order equations of motion), and
\begin{equation}
\Delta = 2 + |m|
\end{equation}
for the operator dual to a first-order two-form, we find
\begin{equation}
\Delta_\phi =4 \; , \; \; \Delta_a =4 \; , \; \; \Delta_u =6 \; , \; \; \Delta_v =8 \; , \; \; \Delta_{\tilde \xi} =3 \; , \; \; \Delta_{\tilde \chi} =7 \; , \quad
\end{equation}
for the operators dual to the scalars,
\begin{equation}
\Delta_{\tilde A_1} =3 \; , \qquad \Delta_{\tilde E_1} =7 \; , \qquad \Delta_{B_1} = \Delta_{C_1} =5 \; ,
\end{equation}
for the operators dual to the vectors, and
\begin{equation}
\Delta_{C_2} = 6 \; , \qquad \Delta_{L_2} =5
\end{equation}
for the operators dual to the two-forms.

These modes should form the bosonic fields of unitary irreducible representations
of $SU(2,2|1)$. The KK modes we have kept are present for any $SE_5$ space and so, in particular,
we can consider the special case $T^{1,1}$ for which the supermultiplet structure was
analysed in detail in \cite{Ceresole:1999zs, Ceresole:1999ht}.
We deduce that the metric and the vector $\tilde A_1$ form a massless graviton multiplet,
the fields $\tilde E_1$, $u$, $v$ and  $\tilde \chi$ fill out a long vector multiplet,
the fields $\xi$, $a$, $\phi$ form a hypermultiplet and
finally, $B_1$, $C_1$, $C_2$ and $L_2$ form a semi-long massive gravitino multiplet.

It is also interesting to consider the special case of $S^5$. The $N=8$ KK spectrum was computed in \cite{Kim:1985ez}
and the modes were arranged in supermultiplets of $SU(2,2|4)$, in \cite{Gunaydin:1984fk}.
The various fields of our $D=5$ theory can be identified with those presented in figures 1, 2 and 3 of \cite{Kim:1985ez}.
Specifically, the metric, the scalars $\phi$, $a$, $\tilde \xi$ and the
vector $\tilde A_1$ belong to the supermultiplet with $p=2$ (following the notation of \cite{Gunaydin:1984fk}),
namely, the $N=8$ $SO(6)$ gauged supergravity multiplet.
Similarly the vectors $B_1$, $C_1$, the two-forms $C_2$, $L_2$ belong to the supermultiplet with $p=3$ and
the scalars $v$, $u$, $\tilde \chi$, the vector $\tilde E_1$ belong to the supermultiplet with $p=4$ (the breathing mode supermultiplet).

\subsubsection{The Romans $AdS_5$ vacuum}
The theory admits another $AdS_5$ vacuum solution where
\be
e^{4U}=e^{-4V}=\frac{2}{3},\quad \xi=\frac{1}{\sqrt{12}}e^{\phi/2}e^{i\theta},\quad\chi-a\xi=ie^{-\phi}\xi
\ee
with arbitrary axion $a$ and dilaton $\phi$ and $\theta$ is an arbitrary constant phase. The $AdS_5$ radius
is $2{\sqrt 2}/3$.
This solution can be uplifted to a class of solutions that were first found by
Romans in \cite{Romans:1984an}, generalising analogous solutions
constructed in $D=11$ supergravity in \cite{Pope:1984bd}\cite{Pope:1984jj}.
For the special case when the $SE_5=S^5$ it is expected that this solution is unstable \cite{hpw}.

\subsection{Further truncations}
There are various additional truncations of the fields appearing in the ansatz \reef{KKmet}, \reef{KKforms}
that are consistent with the type IIB equations of motion. Let us discuss several
of them and in particular make contact with some other works in the literature.
In particular we will recover the truncations of \cite{Maldacena:2008wh} which helped to
motivate the work of \cite{Gauntlett:2009zw} and of this paper.
Note that some cases that we discuss below
can be combined. It is notationally convenient to label some of the forms as $G_i$, $H_i$,
with $i=1,2,3$ and $M_a$, $N_a$ with $a=0,1$.

\subsubsection{Self-dual five-form, dilaton and axion}
It is consistent to truncate IIB supergravity itself to just the ten-dimensional metric, self-dual five-form $F_{(5)}$, dilaton $\Phi$ and axion $C_{(0)}$, by setting $H_{(3)}=F_{(3)} =0$.
It is also consistent to further truncate to just the ten-dimensional metric and self-dual five-form by further setting $\Phi = C_{(0)} =0$.
Accordingly, it is consistent to truncate all of the modes coming from  $H_{(3)}$, $F_{(3)}$ by setting
$H_i=G_i=M_a=N_a=0$. It is also then consistent to further set $\phi =a=0$.

\subsubsection{NS sector}
It is also consistent with the type IIB equations of motion to set all of the Ramond-Ramond fields to zero,
$F_{(5)}=F_{(3)}=F_{(1)}=0$. We can therefore set $e^{Z}=K_1=K_2=L_2=G_i=N_a=a=0$.
Since this would be a universal reduction of type I supergravity on $SE_5$ incorporating the breathing mode,
it seems quite plausible that the resulting theory should be the bosonic part of an $N=2$ gauged supergravity theory.
Indeed the truncated theory contains a metric, two vectors $A_1,B_1$, a two-form $B_2$ and four real scalars $U,V,\phi,b$,
and a complex scalar $\xi$
which could comprise the bosonic part of a gravity multiplet, one vector multiplet, one tensor mutliplet and a single hypermultiplet.
Note that this truncated $D=5$ theory will no longer have an $AdS_5$ vacuum solution.

\subsubsection{No R-charged fields}
It is consistent to set all of the fields carrying non-zero R-charge to zero:
$L_2 =M_a=N_a=0$. Recall that these are the fields appearing with $\Omega$ in \reef{KKforms}.

\subsubsection{Minimal $N=2$ gauged supergravity in $D=5$}
We can recover the KK reduction to minimal $D=5$ $N=2$ gauge supergravity of \cite{Buchel:2006gb}
(see also \cite{Tsikas:1986rx}) by setting  $U=V=e^Z=K_1=L_2=G_i=H_i=M_a=N_a=0$
and $K_2 = - F_2$.  In fact our equations of motion reduce to (2.15), (2.16) of \cite{Buchel:2006gb} with
 $F_2^{\mathrm{here}} = (1/3)F_2^{\mathrm{there}}$.

\subsubsection{The truncations of \cite{Maldacena:2008wh}}
Two consistent truncations of type IIB supergravity on $SE_5$ spaces were studied in \cite{Maldacena:2008wh}, in the context of non-relativistic holography, and both can be
simply obtained from our results.

Firstly, if we set $e^Z=L_2=K_1=K_2=H_3= G_i=M_a=N_a=A_1=a=0$ we obtain the
the truncation discussed in appendix D.1 of \cite{Maldacena:2008wh}.
(We can identify $H_2=F^{\mathrm{there}}_2$ and $H_1=-2A_1^{\mathrm{there}}$.

Secondly, we can also set $e^Z=L_2=G_i=H_i=M_a=N_a=a=\phi=0$ to obtain the truncation discussed
appendix D.2 of \cite{Maldacena:2008wh}. (We can identify
$F_2={\cal F}=d {\cal A}$, $K_1=2{\bf A}$, $K_2=-\bbF=-{\cal F}-d{\bf A}$ to find agreement after taking into account a different convention
for the $D=5$ orientation.)

\subsubsection{Gravity and scalars}
It is also consistent to set $A_1=K_2=K_1=L_2= H_i=G_i=e^Z=M_a=N_a=0$ leaving only the metric and the scalars $U$, $V$, $\phi$, $a$.
It is consistent to then further set $\phi=a=0$ and then $U=V$.
The latter truncation was discussed  in \cite{Bremer:1998zp} \cite{Liu:2000gk} in the context of IIB reductions on $S^5$
(who also considered the addition of some other fields).

\subsubsection{No $p=3$ sector}
At the end of section 3.3.1, for the special case when $SE_5=S^5$, we argued that the modes we have kept arise from
the $p=2,3$ and $p=4$ sectors in the notation of \cite{Gunaydin:1984fk}. Interestingly, for any $SE_5$, it is possible to set all of the fields corresponding to the $p=3$ sector to zero, namely $H_i=G_i=L_2=0$, leaving only the $p=2$ and $p=4$ sectors. Along with the metric, this truncated theory contains five real scalars $U,V,\phi,a,h$, two complex scalars $\xi,\chi$ and two one-forms $A_1,E_1$. It would be interesting to know if this theory is the bosonic part of an $N=2$ gauged supergravity coupled to a
vector multiplet and two hypermultiplets.

Having truncated out the $p=3$ sector for general $SE_5$, it is consistent to further set $a=\phi=0$ while setting one of the scalars to be proportional to the other: $\chi = i  \xi$. (From (\ref{MassesCharged}) we see that this is tantamount to truncating the $p=4$ charged scalar $\tilde \chi$ with mass $m^2_{\tilde \chi} =21$).  Along with the metric, this truncated theory contains three real scalars $U,V,h$, one complex scalar $\xi$ and two one-forms $A_1,E_1$. This theory has a chance to be the bosonic part an $N=2$ gauged supergravity coupled to a vector multiplet and a single hypermultiplet.
Alternatively, having truncated out the $p=3$ sector, it is consistent to further truncate out the $p=4$ sector, leaving only the $p=2$ modes, and again one obtains a theory that is consistent with
being the bosonic part of an $N=2$ gauged supergravity now coupled to a single hypermultiplet.

\subsubsection{The truncation of \cite{Gubser:2009qm}}
For a general $SE_5$, having truncated out the $p=3$ sector ($H_i=G_i=L_2=0$), the dilaton and axion ($a=\phi=0$), and one of the complex scalars ($\chi = i  \xi$), it is consistent to further set
\begin{equation}
e^{4U} = e^{-4V} = 1-4 |\xi|^2
\end{equation}
while also truncating the massive vector $\tilde E_1$ defined in (\ref{MassesVectors}) by setting $K_2 =-  F_2$ and $h=0$.
The resulting theory  contains the metric, a massless vector $A_1$, and a charged scalar $\xi$ with mass $m^2_{\xi} =-3$ in the supersymmetric $AdS_5$ vacuum. In fact we precisely recover the truncation first discussed in \cite{Gubser:2009qm}
in the context of holographic superconductivity (we should set $A^{\textrm{here}} = \tfrac23 A^{\textrm{there}}$, $L^{\textrm{there}}=1$ and   $\xi =  \tfrac12 e^{i\theta} \tanh \tfrac{\eta}{2}$). Note that for the special case when $SE_5=S^5$ the fields kept in this truncation all arise in the $p=2$ sector and hence can be obtained as a truncation of $N=8$ $D=5$ $SO(6)$ gauged supergravity.

This Type IIB truncation has a direct analogue in  $D=11$ supergravity reduced on $SE_7$ which was presented in \cite{Gauntlett:2009bh}
building on \cite{Gauntlett:2009zw}\cite{Gauntlett:2009dn}.

\section{Final Comments}
We conclude with some comments on type IIB reductions for the special case when $SE_5=S^5$ and
then on  $D=11$ reductions on seven-dimensional tri-Sasaki manifolds.

The spectrum of type IIB supergravity on $S^5$ was computed in \cite{Kim:1985ez}
and the modes were arranged in supermultiplets of $SU(2,2|4)$ in \cite{Gunaydin:1984fk}.
We have already noted at the end of section 3.3.1 that the modes that we have kept
in our consistent KK reduction belong to the supermultiplets with $p=2,3$ and $4$ (following the notation of \cite{Gunaydin:1984fk}).
We have also seen in section 3.4.7 that is possible to truncate out the modes arising in the
$p=3$ sector consistently and possibly consistent with $N=2$ supersymmetry.

In \cite{Gauntlett:2009zw} it was conjectured that there might be a consistent truncation of type IIB on
$S^5$ to the full massless graviton multiplet of the $p=2$ sector
combined with the full breathing mode multiplet of the $p=4$ sector and consistent with $N=8$ supersymmetry. The bosonic fields of the $p=2$ multiplet consist of the metric, scalars in the
${\bf 1_C}$, ${\bf 10_C}$, ${\bf 20}$, vectors in the ${\bf 15}$ and a two-form in the ${\bf 6_C}$ of the $SO(6)$ R-symmetry group,
and correspond to the fields of maximal $SO(6)$ gauged supergravity. On the other hand
the $p=4$ multiplet has bosonic field content consisting of a massive graviton in the ${\bf 20}$,
scalars in the ${\bf 105}$, ${\bf 126_C}$, ${\bf 20_C}$, ${\bf 84}$, ${\bf 10_C}$, ${\bf 1}$,
vectors in the ${\bf 175}$, ${\bf 64_C}$, ${\bf 15}$ and two-forms in the ${\bf 50_C}$, ${\bf 45_C}$, ${\bf 6_C}$.
Note that the massive complex two-forms satisfy self-duality equations and hence have six real degrees of freedom
\cite{Townsend:1983xs} and also that the singlet scalar corresponds to the breathing mode.

In light of the results presented in this paper, where for the special case of $SE_5=S^5$ we
included modes in the $p=3$ sector, we might expect that there is a truncation of type IIB on $S^5$ to an $N=8$ theory
that keeps the $p=2,4$ and also the $p=3$ multiplet, whose bosonic content consists of a massive graviton in ${\bf 6}$,
scalars in the ${\bf 50}$, ${\bf 45_C}$, ${\bf 6_C}$, vectors in the ${\bf 64}$, ${\bf 15_C}$
and two-forms in the ${\bf 20_C}$, ${\bf 10_C}$, ${\bf 1_C}$.  Going further one
might conjecture that one could truncate the $p=3$ sector of this conjectured theory to obtain the conjectured theory of \cite{Gauntlett:2009zw}
with the $p=2$ and $p=4$ sectors.
The existence of both massive and massless gravitons combined with the $N=8$ supersymmetry
in these conjectured theories necessarily means that they would have to be very exotic\footnote{While this paper was in press, \cite{Liu:2010ya}  was posted
to the arXiv which contains strong arguments  against these exotic possibilities.}.

Consistent KK reductions of $D=11$ supergravity on
$SE_7$ spaces, corresponding to $AdS_4\times SE_7$ solutions preserving $N=2$ supersymmetry,
%and dual to $N=2$ SCFTs in $d=3$,
were presented in \cite{Gauntlett:2009zw} and it was shown that
the $D=4$ reduced theory also preserves $N=2$ supersymmetry. Similar reductions on manifolds with weak $G_2$
holonomy, $M_7$, corresponding to $AdS_4\times M_7$ solutions preserving $N=1$ supersymmetry,
% and dual to $N=1$ SCFTs in $d=3$,
were also found and it was shown that the $D=4$ reduced theory preserves $N=1$
supersymmetry. It was conjectured in \cite{Gauntlett:2009zw} that the analogous reduction on seven dimensional tri-Sasaki
manifolds, $T_7$,
corresponding to $AdS_4\times T_7$ solutions preserving $N=3$ supersymmetry,
% and
%dual to $N=3$ SCFTs in $d=3$,
would give rise to a $D=4$ reduced theory preserving $N=3$ supersymmetry.
However, in light of the results presented in this paper, we expect that this KK reduction
will give rise to a gauged supergravity theory with $N=4$
supersymmetry\footnote{The possibility that the reduced theory will have $N=4$ supersymmetry was first suggested
in \cite{Billo:2000zs}, using different arguments than we give. In \cite{Billo:2000zs} a different coset for the scalar manifold was suggested than the
one we argue for below.}, with an $AdS_4$ vacuum solution
that will spontaneously partially break the supersymmetry from $N=4$ to $N=3$.
To see this, recall \cite{Boyer:1998sf} that the tri-Sasaki space $T_7$ has a globally defined $SU(2)$ structure, specified
by three two-forms, $J^a$, and three one-forms, $\eta^a$, satisfying $d\eta^a=2(J^a-\epsilon^{abc}\eta^b\wedge\eta^c)$
(locally $T^7$ is an $S^3$ bundle over a four-dimensional quaternionic K\"ahler space).
The supersymmetry and field content of the consistent KK reduction of $D=11$ supergravity on $T_7$ will therefore be the same as
the universal KK reduction of $D=11$ on $HK_4\times T^3$. Hence the consistent KK reduction on $T_7$ should lead to a
$D=4$ $N=4$ gauged supergravity coupled to three vector multiplets. In particular, the scalars should parametrise the coset
$SL(2)/SO(2)\times SO(6,3)/(SO(6)\times SO(3))$.
As in the examples studied in \cite{Gauntlett:2009zw}, there should also be a skew-whiffed version of this
$N=4$ gauged supergravity theory where the basic $AdS_4$ vacuum will break all of the supersymmetry.

\subsection*{Acknowledgements}
We would like to thank Nick Halmagyi, Jim Liu, Jan Louis, Dario Martelli, Carlos N\'u\~nez, Eoin \'O Colg\'ain, Ioannis Papadimitriou, Krzystof Pilch, Paul Smyth, and Nick Warner for helpful discussions. We would also like to thank Seok Kim and Daniel Waldram for helpful
early collaboration. JPG is supported by an EPSRC Senior Fellowship and a Royal Society Wolfson Award. OV is supported by an Alexander von Humboldt postdoctoral fellowship and, partially, through the Spanish Government research grant FIS2008-01980.

\appendix

\section{Type IIB supergravity conventions} \label{appa}

The bosonic sector of IIB supergravity contains the RR forms $F_{(1)}$, $F_{(3)}$, $F_{(5)}$, the NS form $H_{(3)}$, the dilaton $\Phi$ and the metric. The forms satisfy the Bianchi identities
\begin{eqnarray}
&& dF_{(5)} + F_{(3)} \wedge  H_{(3)} =0 \label{F5} \\
&& dF_{(3)}  + F_{(1)} \wedge H_{(3)}=0 \label{F3} \\
&& dF_{(1)} =0 \label{F1} \\
&& dH_{(3)} =0 \label{H3}
\end{eqnarray}
which can be solved by introducing potentials as $F_{(5)} = dC_{(4)} -C_{(2)} \wedge  H_{(3)}$, $F_{(3)} = dC_{(2)} -C_{(0)}  dB_{(2)}$, $F_{(1)} = dC_{(0)}$, $H_{(3)} = dB_{(2)}$.

The equations of motion read:
\begin{eqnarray}
&& *F_{(5)} =F_{(5)} \label{eomF5} \\[8pt]
&& d(e^\Phi *F_{(3)})  - F_{(5)} \wedge H_{(3)}=0 \label{eomF3} \\[8pt]
&& d(e^{2\Phi} *F_{(1)})  +e^{\Phi} H_{(3)} \wedge *F_{(3)} =0 \label{eomF1} \\[8pt]
&& d(e^{-\Phi} *H_{(3)})  -e^{\Phi} F_{(1)} \wedge *F_{(3)} -F_{(3)} \wedge F_{(5)}  =0 \label{eomH3} \\[8pt]
&& d*d\Phi -e^{2\Phi} F_{(1)} \wedge *F_{(1)} +\ft12 e^{-\Phi} H_{(3)} \wedge *H_{(3)} -\ft12 e^{\Phi} F_{(3)} \wedge *F_{(3)} =0 \label{eomPhi}\\[8pt]
&& R_{MN} = \ft12 e^{2\Phi} \nabla_M C_{(0)} \nabla_N C_{(0)} + \ft12 \nabla_M \Phi \nabla_N \Phi
      + \ft{1}{96}F_{MP_1P_2P_3P_4}F_N^{~P_1P_2P_3P_4}  \nonumber \\ && \quad \qquad
      + \ft{1}{4}e^{-\Phi}\left(
         H_M{}^{P_1P_2} H_{NP_1P_2}
         - \ft{1}{12}g_{MN}H^{P_1P_2P_3}H_{P_1P_2P_3} \right) \nonumber \\ && \quad \qquad
      + \ft{1}{4}e^{\Phi}\left(
         F_M{}^{P_1P_2} F_{NP_1P_2}
         - \ft{1}{12}g_{MN}F^{P_1P_2P_3}F_{P_1P_2P_3} \right) . \label{IIBEinstein}
\end{eqnarray}

\section{Details on the KK reduction}\label{appb}
Here we shall provide some details of the KK reduction on $SE_5$. The calculations for the
$HK_4\times S^1$ case are very similar and we omit the details.

We first record some useful algebraic conditions satisfied by the globally defined forms $(J,\Omega,\eta)$ that specify the
$SU(2)$ structure on the $SE_5$ space. We have
$\Omega \wedge \Omega^* = 2J \wedge J$,  $\textrm{vol}(SE_5) = \tfrac12 J \wedge J \wedge \eta$, $*J=J \wedge \eta$, $* \Omega = \Omega \wedge \eta$. We also have $J_{ik} J^{jk} = \delta_i^j$, $\Omega_{ik} \Omega^{jk} = 0$,
$\Omega_{ik} \Omega^{*jk} = 2\delta_i^j -2iJ_i{}^j$ and $J_{ik} \Omega^{jk} = -i\Omega_i{}^j$. In addition,
$J_{[ik} J_{mn]} J^{[jk} J^{mn]} = J_{i[k} J_{mn]} J^{jk} J^{mn} = \ft23 \delta^i_j$.

The KK ansatz for the metric can be written as
\be\label{KKmetnotE}
ds^2_{10}=ds^2_5+e^{2U}ds^2(KE_4)+e^{2V}(\eta+A_1)\otimes(\eta +A_1) \ee
where here $ds^2_5$ is the line element of the external five-dimensional metric. At the end we will convert our results to
the Einstein-frame metric $ds^2_{(E)}$ that we used in the main text. The ansatz for the form field-strengths can be written as
{\setlength\arraycolsep{1pt}
\begin{eqnarray} \label{KKformsnotE}
F_{(5)} &=& 4 e^{-4U-V+Z} \textrm{vol}_5
+e^{-V} * K_2 \wedge J + K_1 \wedge J \wedge J \nonumber \\ && + \left[2e^Z J \wedge J -2e^{-4U +V} *K_1 +K_2 \wedge J \right]\wedge (\eta+A_1)\nn
&&  + \left( e^{-V} *L_2 \wedge \Omega + L_2 \wedge \Omega \wedge (\eta+A_1) +c.c. \right)
\nonumber \\[8pt]
F_{(3)} &=& G_3 +G_2 \wedge (\eta+A_1) +G_1 \wedge J
%+G_0 J \wedge \hat e^5
+\left[ N_1 \wedge \Omega +N_0 \Omega \wedge(\eta+A_1) +c.c. \right]
\nonumber \\[8pt]
 H_{(3)} &=& H_3 +H_2 \wedge (\eta+A_1) +H_1 \wedge J
%+H_0 J \wedge \hat e^5
+\left[ M_1 \wedge \Omega +M_0 \Omega \wedge (\eta+A_1) +c.c. \right]
\nonumber \\[8pt]
C_{(0)} &=& a
\nonumber \\[8pt]
\Phi &=& \phi
\end{eqnarray}
}Here, $\textrm{vol}_5$ and $*$ are the volume form and Hodge dual corresponding to the five-dimensional metric $ds^2_5$ in (\ref{KKmetnotE}).
We use a mostly plus metric convention both in $D=10$ and in $D=5$ and the $D=10$ volume form is given by
$\epsilon_{10}=e^{4U+V}\textrm{vol}_5\wedge \textrm{vol}(SE_5)$.

We now substitute the KK ansatz (\ref{KKmetnotE}), (\ref{KKformsnotE}) into the type IIB Bianchi equations and
equations of motion given in (\ref{F5})--(\ref{IIBEinstein}). We first observe that the ansatz for the five-form has been constructed to be self dual
and thus (\ref{eomF5}) is satisfied.

Equation (\ref{H3}) gives:
\begin{eqnarray}\label{HB}
&& dH_3 +H_2 \wedge F_2 =0 \nonumber \\
&& dH_2 =0 \nonumber \\
&& dH_1 +2H_2 %+H_0 F_2
=0 \nonumber \\
%&& H_0 =0 \nonumber \\
&& DM_1 +M_0 F_2 =0 \nonumber \\
&& DM_0 -3i M_1 =0
\end{eqnarray}
where $DM_1 \equiv dM_1 -3i A_1 \wedge M_1$ and $DM_0 \equiv dM_0 -3i A_1 M_0$.
%We will set $H_0=0$  from now on.

\vspace{1cm}

Equation (\ref{F3}) gives
\begin{eqnarray}\label{F3B}
&& dG_3 +G_2 \wedge F_2 +da \wedge H_3 =0 \nonumber \\
&& dG_2 +da \wedge H_2 =0 \nonumber \\
&& dG_1 +2G_2 +da \wedge H_1
%+G_0 F_2
=0 \nonumber \\
%&& G_0 =0 \nonumber \\
&& DN_1 +N_0 F_2 +da \wedge M_1 =0 \nonumber \\
&& DN_0 -3i N_1 +M_0 da   =0
\end{eqnarray}
where $DN_1 \equiv dN_1 -3i A_1 \wedge N_1$ and $DN_0 \equiv dN_0 -3i A_1 N_0$.
%We will set $G_0=0$  from now on.

\vspace{1cm}

Equation (\ref{F5}) gives:
\begin{eqnarray} \label{eqsfromF5}
&& dK_2 -H_1 \wedge G_2 +H_2 \wedge G_1=0
\nonumber \\[6pt]
&& DL_2 -3i e^{-V} *L_2 -H_3N_0 +M_0G_3 +H_2 \wedge N_1 -M_1 \wedge G_2 =0
\nonumber \\[6pt]
&& dK_1 +2K_2 +2e^Z F_2 -H_1 \wedge G_1 -2M_1 \wedge  N_1^* -2 M_1^* \wedge N_1 =0
\nonumber \\[6pt]
&& de^Z -M_1 N_0^* -M_1^*N_0 +M_0 N_1^* +M_0^* N_1 =0
\nonumber \\[6pt]
&& d(e^{-V} *K_2 ) -4e^{-4U+V} *K_1 +K_2 \wedge F_2 -H_3 \wedge G_1 -H_1 \wedge G_3 =0
\nonumber \\[6pt]
&& D(e^{-V} *L_2) +L_2 \wedge F_2 -H_3 \wedge N_1 -M_1 \wedge G_3 =0
\nonumber \\[6pt]
&& d(e^{-4U+V} *K_1) + \ft12 H_3 \wedge G_2 -\ft12 H_2 \wedge G_3 =0
\end{eqnarray}
where
%$D(e^{-V} *L_2 ) \equiv d(e^{-V} *L_2 ) -3i A_1 \wedge e^{-V} *L_2$ and
$DL_2  \equiv dL_2 -3i A_1 \wedge L_2$

\vspace{1cm}

Equation (\ref{eomF3}) gives:
{\setlength\arraycolsep{0pt}
\begin{eqnarray}
&& d(e^{4U+V+\phi} *G_3) -4e^Z H_3 +2H_2\wedge K_1  -2H_1 \wedge K_2
 -4M_1 \wedge L_2^* -4M_1^* \wedge L_2 \nonumber \\ && \qquad +4e^{-V}M_0 *L_2^*  +4e^{-V} M_0^* *L_2    =0
\nonumber \\[14pt]
&& d(e^{4U-V+\phi} *G_2 )   -4 e^{V+\phi} *G_1 -e^{4U+V+\phi} *G_3 \wedge F_2 +2H_3 \wedge K_1  + 2 e^{-V} H_1 \wedge *K_2 \nonumber \\ && \qquad
 +4e^{-V} M_1 \wedge *L_2^* +4e^{-V} M_1^* \wedge *L_2 =0
\nonumber \\[14pt]
&& d(e^{V+\phi} *G_1) -H_3 \wedge K_2 +e^{-V} H_2 \wedge *K_2  +2 e^{-4U+V} H_1 \wedge * K_1 =0
\nonumber \\[14pt]
&& D(e^{V+\phi} *N_1) -H_3 \wedge L_2 + e^{-V} H_2 \wedge *L_2 +2 e^{-4U+V} M_1 \wedge *K_1  \nonumber \\ && \qquad +e^{-V}  \left(4e^{-4U+Z}M_0 +3i  N_0e^{\phi} \right) \textrm{vol}_5  =0
\end{eqnarray}
}
%where $D(e^V *N_1) \equiv d(e^V *N_1) -3i A_1 \wedge e^V *N_1$.

\vspace{1cm}

Equation (\ref{eomH3}) gives:
{\setlength\arraycolsep{0pt}
\begin{eqnarray}
&& d(e^{4U+V-\phi} *H_3) +4e^Z G_3 -2G_2\wedge K_1  +2G_1 \wedge K_2
 +4N_1 \wedge L_2^* +4N_1^* \wedge L_2 \nonumber \\ && \qquad -4e^{-V}N_0 *L_2^*  -4e^{-V} N_0^* *L_2 -e^{4U+V +\phi} da \wedge *G_3   =0
\nonumber \\[14pt]
&& d(e^{4U-V-\phi} *H_2 )   -4 e^{V-\phi} *H_1 -e^{4U+V-\phi} *H_3 \wedge F_2 -2G_3 \wedge K_1  - 2 e^{-V} G_1 \wedge *K_2 \nonumber \\ && \qquad
 -4e^{-V} N_1 \wedge *L_2^* -4e^{-V} N_1^* \wedge *L_2 -e^{4U-V +\phi}da \wedge *G_2 =0
\nonumber \\[14pt]
&& d(e^{V-\phi} *H_1) +G_3 \wedge K_2 -e^{-V} G_2 \wedge *K_2  -2 e^{-4U+V} G_1 \wedge * K_1 -e^{V+\phi} da \wedge *G_1 =0
\nonumber \\[14pt]
&& D(e^{V-\phi} *M_1) +G_3 \wedge L_2 - e^{-V} G_2 \wedge *L_2 -2 e^{-4U+V} N_1 \wedge *K_1  \nonumber \\ && \qquad -e^{-V}  \left(4e^{-4U+Z}N_0 -3i  M_0e^{-\phi} \right) \textrm{vol}_5 -e^{V+\phi} da \wedge *N_1 =0
\end{eqnarray}
}
%where $D(e^V *N_1) \equiv d(e^V *N_1) -3i A_1 \wedge e^V *N_1$.

\vspace{1cm}

Equation (\ref{eomF1}) gives:
\begin{eqnarray}
&& d(e^{4U+V+2\phi} *da )  + e^{4U+V+\phi} H_3 \wedge *G_3 + e^{4U-V+\phi} H_2 \wedge * G_2 + 2 e^{V+\phi} H_1 \wedge * G_1 \nonumber \\ && \qquad   + 4e^{V+\phi} M_1 \wedge *N_1^* + 4e^{V+\phi} M_1^* \wedge *N_1 + 4e^{-V+\phi} \left(  M_0N_0^* +M_0^*N_0 \right) \textrm{vol}_5 =0 \nonumber \\
\end{eqnarray}

\vspace{1cm}

Equation (\ref{eomPhi}) gives:
\begin{eqnarray}
&& d(e^{4U+V} *d\phi )  - e^{4U+V+2\phi}da \wedge *da +\ft12e^{4U+V-\phi} H_3 \wedge *H_3 -\ft12e^{4U+V+\phi} G_3 \wedge *G_3 \nonumber \\ && \qquad + \ft12 e^{4U-V-\phi} H_2 \wedge * H_2 -\ft12 e^{4U-V+\phi} G_2 \wedge * G_2 +  e^{V-\phi} H_1 \wedge * H_1 -e^{V+\phi} G_1 \wedge * G_1 \nonumber \\ && \qquad   + 4e^{V-\phi} M_1 \wedge *M_1^* -4e^{V+\phi} N_1 \wedge *N_1^* +4e^{-V} \left(  e^{-\phi}|M_0|^2 -e^{\phi} |N_0|^2\right) \textrm{vol}_5 =0 \nonumber \\
\end{eqnarray}

\vspace{1cm}

 Finally, we need to impose the Einstein equation (\ref{IIBEinstein}). To calculate the Ricci tensor we use the orthonormal frame
\begin{eqnarray} \label{KKframe}
&& \bar e^\alpha = e^\alpha \; , \qquad \alpha=0, \ldots , 4 \nonumber \\
&& \bar e^i = e^{U}e^i \; , \qquad i=1, \ldots , 4 \nonumber \\
&& \bar e^5 = e^V \hat e^5 \equiv e^V ( \eta + A_1) .
\end{eqnarray}
We find the spin connection is given by
\begin{eqnarray}
&& \bar \omega^{\alpha \beta} = \omega^{\alpha \beta} -\tfrac{1}{2}e^{2V}F^{\alpha \beta} \hat e^5 \nn
&& \bar \omega^{\alpha i} = -e^U \partial^\alpha U e^i \nn
&& \bar \omega^{\alpha 5} = -e^V \partial^\alpha V \hat  e^5 -\tfrac{1}{2}e^{V}F^{\alpha}{}_{ \beta} e^\beta \nn
&& \bar \omega^{ij} = \omega^{ij} -e^{2V-2U} J^{ij} \hat e^5 \nn
&& \bar \omega^{i5} = -e^{V-U} J^i{}_je^j
\end{eqnarray}
and the Riemann tensor, $\bar \Theta^{AB}= d\bar \omega^{AB} + \bar \omega^A{}_C \wedge \bar \omega^{CB}$, by:
\begin{eqnarray}
\bar \Theta^{\alpha \beta} &=& \Theta^{\alpha \beta} -\tfrac14 e^{2V} \left[ F^{\alpha \beta} F_{\lambda \mu} + F^{\alpha}{}_{[\lambda} F^{\beta}{}_{\mu]} \right] \bar e^{\lambda \mu} - \left[ \tfrac12 e^{-V} \nabla_\lambda (e^{2V} F^{\alpha \beta} )  +e^V\left(F^{[\alpha}{}_\lambda \nabla^{\beta]}V \right) \right] \bar e^{\lambda 5} \nonumber \\[10pt]
\bar \Theta^{\alpha i} &=& - \left[ (\nabla_\lambda \nabla^\alpha U +\nabla_\lambda U \nabla^\alpha U )\delta^i_j + \ft12 e^{-2U+2V} J^i{}_j F^\alpha{}_\lambda \right] \bar e^{\lambda j} \nonumber \\
&& +\left[e^{-2U+V} \nabla^\alpha (V-U) J^i{}_j -\ft12 e^V F^{\alpha \gamma} \nabla_\gamma U \delta^i_j \right] \bar e^{j5} \nonumber \\[10pt]
\bar \Theta^{\alpha 5} &=& -\ft12 \left[\nabla_\lambda (e^V F^\alpha{}_\mu ) + e^V\nabla^\alpha V F_{\lambda \mu} \right] \bar e^{\lambda \mu} \nonumber - \left[ \nabla_\lambda \nabla^\alpha V +\nabla_\lambda V \nabla^\alpha V + \ft14 e^{2V} F^{\alpha \gamma} F_{\gamma \lambda} \right] \bar e^{\lambda 5}  \nonumber \\ && -e^{-2U+V} \nabla^\alpha (V-U) J_{ij} \bar e^{ij} \nonumber \\[10pt]
\bar \Theta^{ij} &=& \Theta^{ij} -\ft12 e^{-2U+2V} F_{\alpha \beta} J^{ij} \bar e^{\alpha \beta} -e^{-2U+V} \nabla_\alpha (V-U) J^{ij} \bar e^{\alpha 5} \nonumber \\
&& -\left[ e^{-4U+2V} (J^{ij}J_{hk} +J^i{}_{[h} J^j{}_{k]} ) +\nabla_\gamma U \nabla^\gamma U \delta^i_{[h} \delta^j_{k]} \right] \bar e^{hk} -e^{-3U+V} \nabla_k J^{ij} \bar e^{k 5} \nonumber \\[10pt]
\bar \Theta^{i5} &=& \left[-e^{-2U+V} \nabla_\alpha (V-U) J^i{}_j +\ft12 e^V F_{\alpha\gamma} \nabla^\gamma U \delta^i_j \right] \bar e^{\alpha j} +\left[ e^{-4U+2V} - \nabla_\gamma U \nabla^\gamma V \right] \delta^i_j \bar e^{j 5} \nonumber \\
\end{eqnarray}
Finally the Ricci tensor, $\bar R^A{}_B = \bar \Theta^{AC}{}_{BC}$, is given by
\begin{eqnarray}
&& \bar R_{\alpha \beta} = R_{\alpha \beta} -4 \left(\nabla_\beta \nabla_\alpha U + \partial_\alpha U \partial_\beta U \right) -\left(\nabla_\beta \nabla_\alpha V + \partial_\alpha V \partial_\beta V \right) -\tfrac{1}{2}e^{2V} F_{\alpha \gamma} F_\beta{}^\gamma \nn
&& \bar R_{\alpha i} = 0 \nn
&& \bar R_{\alpha 5} = -\tfrac{1}{2} e^{-2V-4U} \nabla_\gamma \left(e^{3V+4U} F^{\gamma \alpha} \right) \nn
&& \bar R_{ij} = \delta_{ij} \left[6e^{-2U} -2e^{2V-4U} -\nabla_\gamma \nabla^\gamma U -4\partial_\gamma U \partial^\gamma U -\partial_\gamma U \partial^\gamma V \right] \nn
&& \bar R_{i5} = 0 \nn
&& \bar R_{55} = 4e^{2V-4U} -\nabla_\gamma \nabla^\gamma V -4\partial_\gamma U \partial^\gamma V -\partial_\gamma V \partial^\gamma V +\tfrac{1}{4}e^{2V} F_{\alpha \beta} F^{\alpha\beta}
\end{eqnarray}

 The Einstein equations (\ref{IIBEinstein}) now reduce to the following four equations in $D=5$:
\begin{eqnarray}
\label{Ein1} R_{\alpha \beta} &=& 4 \left(\nabla_\beta \nabla_\alpha U + \partial_\alpha U \partial_\beta U \right) +\left(\nabla_\beta \nabla_\alpha V + \partial_\alpha V \partial_\beta V \right) +\ft12 e^{2\phi} \partial_\alpha a \partial_\beta a + \ft12 \partial_\alpha \phi \partial_\beta \phi  \nonumber \\
&& -e^{-4U-2V}\left( 4e^{-4U+2Z} +e^{-\phi} |M_0|^2 +e^{\phi} |N_0|^2 \right) \eta_{\alpha \beta} \nonumber  \\
&& +2 e^{-8U} \left(K_\alpha K_\beta -\ft12 \eta_{\alpha \beta} K_\lambda K^\lambda \right)  + e^{-4U-2V} \left(K_{\alpha \lambda} K_\beta{}^\lambda -\ft14 \eta_{\alpha \beta} K_{\lambda \mu} K^{\lambda \mu} \right) \nonumber \\
&& +\tfrac{1}{2}e^{2V} F_{\alpha \gamma} F_\beta{}^\gamma +4e^{-4U-2V} \left(-L_{ \lambda (\alpha} L^*_{\beta)}{}^\lambda   -\ft14 \eta_{\alpha \beta} L^*_{\lambda \mu} L^{\lambda \mu} \right) \nonumber \\
&& +\ft14 e^{-\phi} \left( H_{\alpha \lambda \mu} H_\beta{}^{\lambda \mu}  -\ft{1}{12} \eta_{\alpha \beta} H_{\lambda \mu \nu} H^{\lambda \mu \nu} \right) +\ft12 e^{-2V-\phi}\left( H_{\alpha \lambda} H_\beta{}^{\lambda} -\ft{1}{8} \eta_{\alpha \beta} H_{\lambda \mu} H^{\lambda \mu} \right) \nonumber \\
&& +e^{-4U-\phi}\left( H_{\alpha} H_\beta   -\ft{1}{4} \eta_{\alpha \beta} H_{\lambda} H^{\lambda } \right) +4e^{-4U-\phi}\left( M_{(\alpha} M^*_{\beta)} -\ft{1}{4} \eta_{\alpha \beta} M^*_{\lambda} M^{\lambda } \right) \nonumber \\
&& +\ft14 e^{\phi} \left( G_{\alpha \lambda \mu} G_\beta{}^{\lambda \mu}  -\ft{1}{12} \eta_{\alpha \beta} G_{\lambda \mu \nu} G^{\lambda \mu \nu} \right) +\ft12 e^{-2V+\phi}\left( G_{\alpha \lambda} G_\beta{}^{\lambda} -\ft{1}{8} \eta_{\alpha \beta} G_{\lambda \mu} G^{\lambda \mu} \right) \nonumber \\
&& +e^{-4U+\phi}\left( G_{\alpha} G_\beta   -\ft{1}{4} \eta_{\alpha \beta} G_{\lambda} G^{\lambda } \right) +4e^{-4U+\phi}\left( N_{(\alpha} N^*_{\beta)} -\ft{1}{4} \eta_{\alpha \beta} N^*_{\lambda} N^{\lambda } \right)
\end{eqnarray}

\begin{eqnarray}
&& d(e^{4U+3V} *F_2) +K_2 \wedge K_2 +4L_2 \wedge L_2^* -8e^{-4U+V+Z} *K_1 -e^{4U+V-\phi} H_2 \wedge *H_3 \nonumber \\ && \qquad -e^{4U+V+\phi} G_2 \wedge *G_3  -4e^{V-\phi}(M_0^* *M_1 + M_0 *M_1^*)  \nonumber \\ && \qquad
-4e^{V+\phi}(N_0^* *N_1 + N_0 *N_1^*) =0
\end{eqnarray}

\begin{eqnarray}  \label{Ein3form}
&& d(e^{4U+V} *dU ) +e^{-4U+V}K_1\wedge *K_1 -\ft18 e^{4U+V-\phi} H_3 \wedge *H_3 -\ft18 e^{4U+V+\phi} G_3 \wedge *G_3 \nonumber \\ && \qquad - \ft18 e^{4U-V-\phi} H_2 \wedge * H_2  -\ft18 e^{4U-V+\phi} G_2 \wedge * G_2 +  \ft14 e^{V-\phi} H_1 \wedge * H_1 +\ft14  e^{V+\phi} G_1 \wedge * G_1  \nonumber \\ && \qquad + e^{V-\phi} M_1 \wedge *M_1^*  +e^{V+\phi} N_1 \wedge *N_1^* \nonumber \\ && \qquad + \left( -6e^{2U+V} +2e^{3V} +  4e^{-4U-V+2Z} +  e^{-V-\phi}|M_0|^2 +e^{-V+\phi} |N_0|^2\right) \textrm{vol}_5 =0 \nonumber \\
\end{eqnarray}

\begin{eqnarray} \label{Ein4form} && d(e^{4U+V} *dV )  -\ft18 e^{4U+V-\phi} H_3 \wedge *H_3 -\ft18 e^{4U+V+\phi} G_3 \wedge *G_3 - \ft12 e^{4U+3V} F_2 \wedge * F_2 \nonumber \\ && \qquad   -e^{-4U+V} K_1 \wedge *K_1 +\ft12 e^{-V} K_2 \wedge *K_2  +2 e^{-V} L_2 \wedge *L_2^*  + \ft38 e^{4U-V-\phi} H_2 \wedge * H_2 \nonumber \\ && \qquad +\ft38 e^{4U-V+\phi} G_2 \wedge * G_2 -  \ft14 e^{V-\phi} H_1 \wedge * H_1 -\ft14  e^{V+\phi} G_1 \wedge * G_1   - e^{V-\phi} M_1 \wedge *M_1^* \nonumber \\ && \qquad  -e^{V+\phi} N_1 \wedge *N_1^* + \left( -4e^{3V} +4e^{-4U-V+2Z} +  3 e^{-V-\phi}|M_0|^2 +3e^{-V+\phi} |N_0|^2\right) \textrm{vol}_5 =0 \nonumber \\
\end{eqnarray}

All the dependence on the internal SE$_5$ has dropped out from the type IIB equations of motion.
This proves the consistency of the KK ansatz (\ref{KKmetnotE}), (\ref{KKformsnotE}). The Lagrangian that gives rise to the above equations of motion is
given by
\begin{equation} \label{lagrangianse2}
{\cal L} = {\cal L}_{\textrm{kin}} + {\cal L}_{\textrm{pot}} + {\cal L}_{\textrm{top}}
\end{equation}
with
\begin{eqnarray} \label{kinetic}
{\cal L}_{\textrm{kin}} &=& e^{4U+V} R \ \textrm{vol}_5 + e^{4U+V} \left(12 dU \wedge *dU + 8 dU \wedge * dV \right) -\ft12 e^{4U+V+2\phi} da \wedge * da \nn
&&-\ft12 e^{4U+V} d\phi \wedge * d\phi
-4e^{V-\phi} M_1 \wedge * M^*_1 -4e^{V+\phi} N_1 \wedge * N^*_1  -2 e^{-4U+V} K_1 \wedge * K_1\nn&&
- e^{V-\phi} H_1 \wedge * H_1 - e^{V+\phi} G_1 \wedge * G_1-\ft12 e^{4U+3V} F_2\wedge * F_2\nn&&
 - e^{-V} K_2 \wedge *K_2 -4 e^{-V} L_2 \wedge * L^*_2
 -\ft{1}{2} e^{4U-V-\phi} H_2 \wedge * H_2\nn&& -\ft{1}{2} e^{4U-V+\phi} G_2 \wedge * G_2
 -\ft{1}{2} e^{4U+V-\phi} H_3 \wedge * H_3 -\ft{1}{2} e^{4U+V+\phi} G_3 \wedge * G_3
\end{eqnarray}
and
\begin{eqnarray} \label{potential}
{\cal L}_{\textrm{pot}} &=& \Big[ 24e^{2U+V} -4e^{3V} -8e^{-4U-V} \big(1 +\ft{i}{3}(M_0^*N_0 -M_0N_0^*) \big)^2 \nonumber \\ && \; -4e^{-V-\phi}|M_0|^2 -4e^{-V+\phi}|N_0|^2 \Big] \textrm{vol}_5\nn
&=& \Big[ 24e^{2U+V} -4e^{3V}
-8e^{-4U-V} \big(1  +3i(\xi^*\chi-\xi\chi^*) \big)^2 \nonumber \\ && \;
-36e^{-V-\phi}|\xi|^2 -36e^{-V+\phi}|\chi-a\xi|^2 \Big] \textrm{vol}_5
 \end{eqnarray}
 and ${\cal L}_{\textrm{top}} $ is given in \reef{ltopmmm}.
The Einstein frame Lagrangian can be obtained by the change of metric $g_{\mu \nu}^{(E)} = e^{\frac{2}{3}(4U+V)}g_{\mu \nu}$, to obtain
\begin{equation} \label{lagEinframe}
{\cal L}^{(E)} = {\cal L}_{\textrm{kin}}^{(E)} + {\cal L}_{\textrm{pot}}^{(E)} + {\cal L}_{\textrm{top}} \; ,
\end{equation}
where ${\cal L}_{\textrm{kin}}^{(E)} $ and ${\cal L}_{\textrm{pot}}^{(E)} $ are given in \reef{kineticEinframeSE} and \reef{potentialEinframe},
respectively, and ${\cal L}_{\textrm{top}}$ is unchanged.

\end{document}